\documentclass[sigconf, screen]{acmart}

\setcopyright{none}
\renewcommand\footnotetextcopyrightpermission[1]{}

\AtBeginDocument{%
  \providecommand\BibTeX{{%
    Bib\TeX}}}

\setcopyright{none} 
\copyrightyear{2025}
\acmYear{2025}
\acmDOI{XXXXXXX.XXXXXXX}

\acmConference[MICRO 2026]{The 59th IEEE/ACM International Symposium on Microarchitecture}{October 31-- Nov 4, 2026}{Athens, Greece}

\acmISBN{978-X-XXXX-XXXX-X/XX/XX}

\usepackage{multirow}
\usepackage{comment}
\usepackage{verbatim} 
\usepackage{tikz}
\usepackage{pgfplots}
\usepackage{pgfplotstable}
\usepackage{graphicx}
\usepackage{pifont}
\usepackage[export]{adjustbox}
\usepackage{pgf-pie}
\usepackage{tkz-graph}
\usepackage{graphicx}
\usepackage{subcaption}
\usepackage[skip=2pt]{caption}
\usepackage{hyperref}
\usepackage{microtype}
\usepackage[ruled,linesnumbered,vlined]{algorithm2e}
\usepackage{amsmath}
\usepackage{xcolor} 
\usepackage{listings}
\usepackage{tcolorbox}
\usepackage{tabularx} 
\usepackage{booktabs}

\definecolor{SlateBlue}{RGB}{106, 90, 205} 
\definecolor{CadetBlue}{RGB}{95, 158, 160} 
\definecolor{BurntOrange}{RGB}{204, 85, 0} 
\definecolor{Goldenrod}{RGB}{218, 165, 32} 

\hypersetup{
    colorlinks,
    linkcolor={red!40!black},
    citecolor={blue!40!black},
    urlcolor={black}
}
\usepackage{booktabs}

\usetikzlibrary{pgfplots.groupplots}
\usetikzlibrary{shadows,patterns,shapes,arrows,decorations.pathmorphing,backgrounds,positioning,fit,plotmarks,calc,spy,matrix}
\pgfplotsset{compat=1.11,
    /pgfplots/ybar legend/.style={
    /pgfplots/legend image code/.code={%
       \draw[##1,/tikz/.cd,yshift=-0.25em]
        (0cm,0cm) rectangle (3pt,0.8em);},
   },
}
\usetikzlibrary{decorations.text}
\definecolor{boxbg}{rgb}{240,240,240}
\definecolor{cadetblue}{rgb}{0.37, 0.62, 0.63}
\definecolor{airforceblue}{rgb}{0.36, 0.54, 0.66}
\definecolor{caribbeangreen}{rgb}{0.0, 0.8, 0.6}
\definecolor{carolinablue}{rgb}{0.6, 0.73, 0.89}
\definecolor{darkgoldenrod}{rgb}{0.72, 0.53, 0.04}
\definecolor{debianred}{rgb}{0.84, 0.04, 0.33}
\definecolor{fuzzywuzzy}{rgb}{0.8, 0.4, 0.4}
\definecolor{grullo}{rgb}{0.66, 0.6, 0.53}
\definecolor{ceil}{rgb}{0.57, 0.63, 0.81}
\definecolor{candypink}{rgb}{0.89, 0.44, 0.48}
\definecolor{calpolypomonagreen}{rgb}{0.12, 0.3, 0.17}
\definecolor{burntsienna}{rgb}{0.91, 0.45, 0.32}
\definecolor{atomictangerine}{rgb}{1.0, 0.6, 0.4}
\definecolor{goldenrod}{rgb}{0.85, 0.65, 0.13}
\definecolor{gamboge}{rgb}{0.89, 0.61, 0.06}
\definecolor{amber}{rgb}{1.0, 0.75, 0.0}
\definecolor{battleshipgrey}{rgb}{0.52, 0.52, 0.51}
\definecolor{darkcerulean}{rgb}{0.03, 0.27, 0.49}
\definecolor{fuzzywuzzy}{rgb}{0.8, 0.4, 0.4}
\definecolor{mediumseagreen}{rgb}{0.24, 0.7, 0.44}
\definecolor{antiquebrass}{rgb}{0.8, 0.58, 0.46}
\definecolor{apricot}{rgb}{0.98, 0.81, 0.69}
\definecolor{asparagus}{rgb}{0.53, 0.66, 0.42}
\definecolor{bananamania}{rgb}{0.98, 0.91, 0.71}
\definecolor{cadmiumgreen}{rgb}{0.0, 0.42, 0.24}
\definecolor{chocolate}{rgb}{0.48, 0.25, 0.0}
\definecolor{cinereous}{rgb}{0.6, 0.51, 0.48}
\definecolor{aliceblue}{rgb}{0.94, 0.97, 1.0}
\definecolor{beaublue}{rgb}{0.74, 0.83, 0.9}
\definecolor{blizzardblue}{rgb}{0.67, 0.9, 0.93}
\definecolor{bittersweet}{rgb}{1.0, 0.44, 0.37}
\definecolor{camouflagegreen}{rgb}{0.47, 0.53, 0.42}
\definecolor{darkolivegreen}{rgb}{0.33, 0.42, 0.18}
\definecolor{darkpastelblue}{rgb}{0.47, 0.62, 0.8}
\definecolor{desertsand}{rgb}{0.93, 0.79, 0.69}
\definecolor{deeppeach}{rgb}{1.0, 0.8, 0.64}
\definecolor{indianred}{rgb}{0.8, 0.36, 0.36}
\definecolor{oldmauve}{rgb}{0.4, 0.19, 0.28}
\definecolor{lightblue}{rgb}{0.68, 0.85, 0.9}
\definecolor{lightcyan}{rgb}{0.88, 1.0, 1.0}
\definecolor{viridian}{rgb}{0.25, 0.51, 0.43}
\definecolor{slategray}{rgb}{0.44, 0.5, 0.56}
\definecolor{manatee}{rgb}{0.59, 0.6, 0.67}
\definecolor{darkbrown}{rgb}{0.4, 0.26, 0.13}
\definecolor{almond}{rgb}{0.94, 0.87, 0.8}
\def\BibTeX{{\rm B\kern-.05em{\sc i\kern-.025em b}\kern-.08em
    T\kern-.1667em\lower.7ex\hbox{E}\kern-.125emX}}
\usepackage{colortbl}
\usepackage{tabulary}
\pgfkeys{%
/piechartthreed/.cd,
scale/.code                =  {\def\piechartthreedscale{#1}},
mix color/.code            =  {\def\piechartthreedmixcolor{#1}},
background color/.code     =  {\def\piechartthreedbackcolor{#1}},
name/.code                 =  {\def\piechartthreedname{#1}}}
\newcommand\piechartthreed[2][]{%
   \pgfkeys{/piechartthreed/.cd,
     scale            = 1,
     mix color        = gray,
     background color = white,
     name             = pc} 
  \pgfqkeys{/piechartthreed}{#1}
  \begin{scope}[scale=\piechartthreedscale] 
  \begin{scope}[xscale=5,yscale=3] 
     \path[preaction={fill=black,opacity=.8,
         path fading=circle with fuzzy edge 20 percent,
         transform canvas={yshift=-15mm*\piechartthreedscale}}] (0,0) circle (1cm);
     \pgfmathsetmacro\totan{0} 
     \global\let\totan\totan 
     \pgfmathsetmacro\bottoman{180} \global\let\bottoman\bottoman 
     \pgfmathsetmacro\toptoman{0}   \global\let\toptoman\toptoman 
     \begin{scope}[draw=black,thin]
     \foreach \an/\col [count=\xi] in {#2}{%
     \def\space{ } 
        \coordinate (\piechartthreedname\space\xi) at (\totan+\an/2:0.75cm); 
        \ifdim 180pt>\totan pt 
         \ifdim 0pt=\toptoman pt
            \pgfmathsetmacro\toptoman{180} 
            \global\let\toptoman\toptoman         
            \else
          \fi
        \fi   
        \fill[\col!80!gray,draw=black] (0,0)--(\totan:1cm)  arc(\totan:\totan+\an:1cm)
                                     --cycle;     
       \pgfmathsetmacro\finan{\totan+\an}
       \ifdim 180pt<\finan pt 
         \ifdim 180pt=\bottoman pt
            \shadedraw[left color=\col!20!\piechartthreedmixcolor,
                       right color=\col!5!\piechartthreedmixcolor,
                       draw=black,very thin] (180:1cm) -- ++(0,-3mm) arc (180:\totan+\an:1cm) 
                                                       -- ++(0,3mm)  arc (\totan+\an:180:1cm);
            \pgfmathsetmacro\bottoman{0}
            \global\let\bottoman\bottoman
            \else
            \shadedraw[left color=\col!20!\piechartthreedmixcolor,
                       right color=\col!5!\piechartthreedmixcolor,
                       draw=black,very thin](\totan:1cm)-- ++(0,-3mm) arc(\totan:\totan+\an:1cm)
                                                        -- ++(0,3mm)  arc(\totan+\an:\totan:1cm); 
          \fi
        \fi
        \pgfmathsetmacro\totan{\totan+\an}  \global\let\totan\totan 
       } 
    \end{scope}
   \end{scope}  
\end{scope}
}

\definecolor{goodgreen}{HTML}{009E73}
\definecolor{badred}{HTML}{D55E00}

\newcommand{\cmark}{\textcolor{goodgreen}{\large\ding{51}}}
\newcommand{\xmark}{\textcolor{badred}{\large\ding{55}}}

\lstdefinelanguage{pseudocode}{
  keywords={function, return, if, then, else, while, for, to, do, break, true, false},
  morecomment=[l]{//},
  morestring=[b]",
}

\begin{document}

\newcommand\name{\textit{COMPOSE}}
\title{\name: Static Timing-driven Composable Reconfigurable Architecture \textcolor{black}{for Accelerating Recurrence-Bound Loops}}

\author{Rohan Juneja}
\email{rohan@comp.nus.edu.sg}
\affiliation{
    \department{School of Computing}
    \institution{National University of Singapore}
    \country{}
}

\author{Vishruti Ranjan}
\email{vishruti@comp.nus.edu.sg}
\affiliation{
    \department{School of Computing}
    \institution{National University of Singapore}
    \country{}
}

\author{Rakshith Harish}
\email{Rakshith_Harish@a-star.edu.sg}
\affiliation{
    \department{Institute of Microelectronics}
    \institution{A*STAR}
    \country{}
}

\author{Li-Shiuan Peh}
\email{peh@comp.nus.edu.sg}
\affiliation{
    \department{School of Computing}
    \institution{National University of Singapore}
    \country{}
}




\begin{abstract}
Coarse-Grained Reconfigurable Architectures (CGRAs) provide a spatially programmable substrate well suited for accelerating compute-intensive workloads with abundant parallelism. However, traditional CGRA execution models rely on rigid, fixed-size processing elements (PEs) that are statically bound to individual operations, which forces inter-iteration dependencies to be resolved through serialized scheduling. This limits throughput and reduces parallelism across loop iterations. Moreover, static execution schedules often fail to exploit available timing slack between operations, leading to resource underutilization and increased latency. The frequent registering of intermediate results further exacerbates pressure on register files and local memories, introducing data movement overheads that reduce energy efficiency, particularly in power or memory constrained environments.

To address these challenges, we introduce {\name}, a composable CGRA architecture that enables dynamic formation of PEs at compile time guided by static timing information. By spatially fusing operations across loop iterations and selectively utilizing slack, {\name} resolves inter-iteration dependencies that limit throughput and enables low latency execution by reducing slack wastage. Additionally, the architecture reduces register file pressure by deferring output registration when intermediate values remain locally consumable, which significantly lowers redundant memory traffic. Across a diverse set of workloads, {\name} on average delivers 1.6x performance improvement and 2.9x EDP reduction over state-of-the-art (SOTA), at minimal area and power overheads.
\end{abstract}


\maketitle
\section{Introduction}

\textbf{CGRA Design Space.} 
Coarse-grained reconfigurable architectures (CGRAs) bridge the gap between fine-grained FPGA lookup table fabrics and fixed-function ASIC accelerators by providing a spatially programmable array of processing elements (PEs) connected by a reconfigurable interconnect. Both the industry~\cite{sambanova_sn40l, intel_csa, renesas_drp, samsung_srp} and academia~\cite{adres, agile, snafu, amber, nexus, canon} have built many variants that differ in array size, interconnect structure, and mix of compute and memory tiles, all with the goal of mapping compute kernels onto a regular fabric for high throughput and good energy efficiency.

These architectures are mapped with a software toolchain that starts from a kernel and builds a dataflow graph (DFG) where nodes are operations and edges are data dependencies. The mapper targets throughput via the initiation interval (II), the number of cycles between the start of consecutive loop iterations. Smaller II means higher steady-state throughput. It then selects an unrolling factor to expose parallelism, places operations on PEs, and routes dependencies through the interconnect under resource and connectivity constraints. Finally, the toolchain builds a schedule, and outputs the configuration for PEs and switches in the interconnect.

\begin{figure}[t!]
    \centering
    \hspace{-0.5cm}
    \includegraphics[width=\columnwidth]{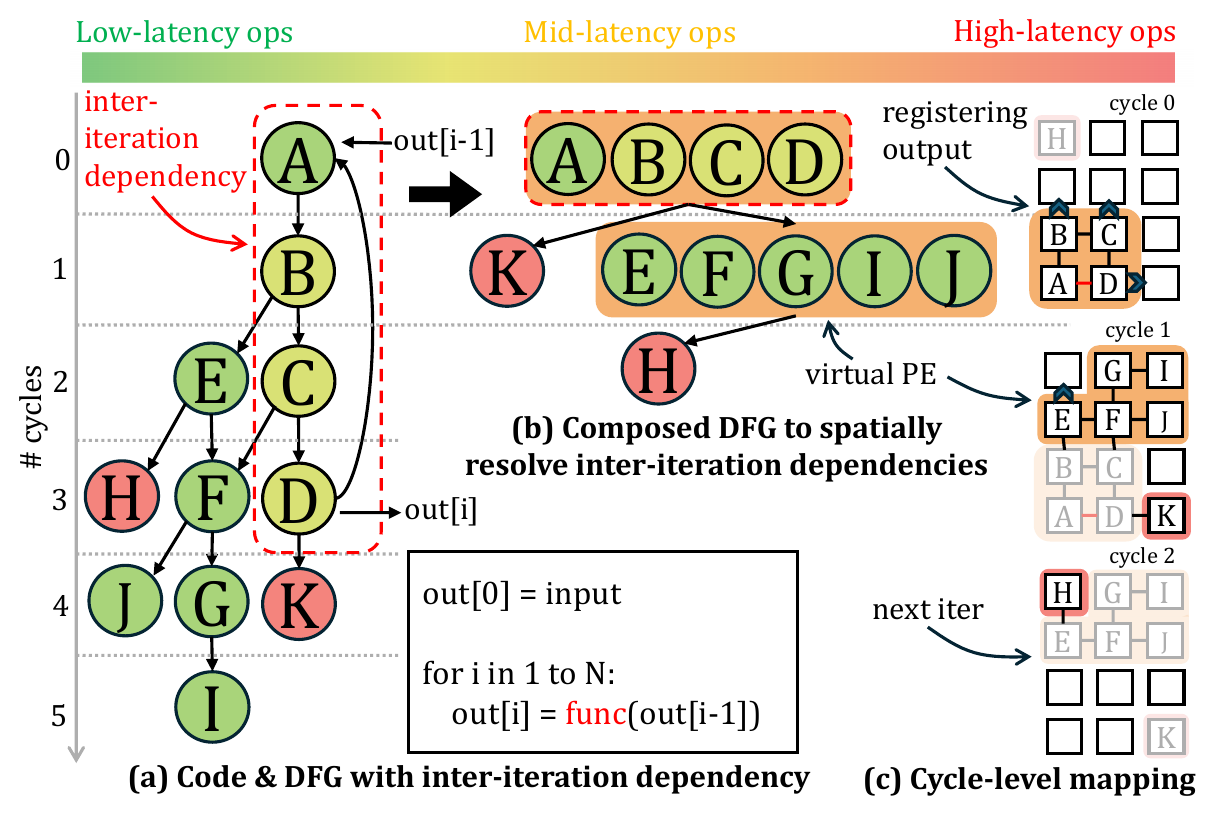}
    \caption{Timing-aware spatial composition in {\name}. Fusing low-latency operations into virtual PEs absorbs unexploited timing slack, compressing the DFG depth and resolving inter-iteration dependencies without violating the global clock period.}
    \label{fig:compose_intro}
\end{figure}


\textbf{Revisiting PE Boundaries in CGRAs.}
Prior CGRA research lacks a unified abstraction for the Processing Element (PE). While some define a PE as a fixed-function unit and others as a complete ALU-based compute node, we adopt a more precise definition: the PE is the atomic scheduling unit to which exactly one operation is mapped per clock cycle. Under this definition, conventional CGRA compilers face significant inefficiencies when mapping complex or composite instructions. Standard flows decompose high-level operations into primitive instructions, each partitioned by a rigid cycle boundary that necessitates a write-back and subsequent read from a local register file. These boundaries waste timing slack, as the global clock period is governed by the critical path of the worst-case operation. For inter-iteration data dependencies, this forces unnecessary serialization, thereby increasing execution latency and degrading effective throughput.

This inefficiency is further exacerbated in power-constrained domains where architectures are operated at reduced frequencies. In edge devices, for instance, designers often down-clock high-frequency fabrics to minimize dynamic power, resulting in substantial unexploited timing slack.

\textbf{Proposed Composable PEs.}
These inefficiencies arise because conventional CGRA compilation flows remain architecturally agnostic to fine-grained circuit-level timing, assigning a uniform clock cycle budget to all operations regardless of their physical delay. In practice, critical-path-delays exhibit significant variance; while high-latency units define the global clock period, low- and mid-latency operations execute much faster, leaving substantial unexploited timing slack. As illustrated in Fig.~\ref{fig:compose_intro}(a), the mapping of our motivational DFG onto a conventional fabric assigns each operation to a discrete cycle (spanning 6 cycles in total), ignoring the potential to exploit this slack.

To address this, we introduce {\name}, a timing-aware framework that leverages static timing analysis to dynamically establish ``virtual PE'' boundaries constrained by the target clock period and physical resource availability. As shown in Fig.~\ref{fig:compose_intro}(b) and (c), {\name} harvests the slack inherent in high-speed operations to spatially chain dependent producer-consumer pairs (e.g., grouping nodes A, B, C, and D) within a single clock cycle, without violating the timing constraints set by the slowest path. This spatial composition is realized through a low-latency hardware bypassing mechanism that enables direct combinational chaining across PEs, effectively bypassing the standard clocked register boundaries (detailed in Section~\ref{sec:design}) and only registering the output at the boundary of the virtual PE.

This provides 3 key advantages: \textcolor{black}{(1) it enables sub-cycle recurrence resolution by composing PEs such that critical inter-iteration dependencies complete within fewer cycles (e.g., the loop-carried dependency from ($D_{out[i]}$ to $A_{out[i-1]}$),} (2) it collapses the DFG depth by fusing combinational chains to minimize input-to-output latency (reducing execution from 6 cycles to 3), and (3) it reduces register-file pressure and dynamic power by achieving equivalent or superior performance at a reduced clock frequency, effectively trading unexploited timing slack for significant energy savings.

\begin{table}[h]
\centering
\scriptsize
\setlength{\tabcolsep}{4pt}
\renewcommand{\arraystretch}{1.12}

\resizebox{\columnwidth}{!}{%
\begin{tabular}{lccc}
\toprule
\textbf{Related Works} &
\textbf{PE composition}&\textbf{Compilation} &
\textbf{Inter-iter.} \\
\textbf{} &
\textbf{decided at}&\textbf{Slack-aware?} &
\textbf{deps. Aware?} \\
\midrule

EGRA~\cite{egra_tvlsi,egra_sasp} &  Fabrication & \xmark & \xmark \\

APEX~\cite{apex_asplos2023} &  Fabrication & \cmark & \xmark \\

Slack-aware sched.~\cite{slack_aware} &  Compilation & \cmark & \xmark \\

CGRA-Express~\cite{cgra_express} &  Compilation & \cmark & \xmark \\

Cascade~\cite{cascade} & Compilation & \cmark & \xmark \\

Recurrence-aware sched.~\cite{recurrence_aware_sigplan2009} &  \xmark & \xmark & \cmark \\

UE-CGRA~\cite{uecgra} &  \xmark & \xmark & \cmark \\


\specialrule{0.4pt}{2pt}{0pt}
\rowcolor[HTML]{C9C0BB}
\rule{0pt}{3.5ex}\textbf{{\name}} &
\rule{0pt}{3.5ex}\textbf{Compilation} &
\rule{0pt}{3.5ex}\textbf{\cmark} &
\rule{0pt}{3.5ex}\textbf{\cmark} \\
\specialrule{0.8pt}{0pt}{0pt}

\end{tabular}%
}

\caption{\textcolor{black}{Comparison with prior CGRA composition and scheduling approaches. {\name} is the only compilation-time approach that is both slack-aware and explicitly accounts for inter-iteration dependencies.}}
\label{tab:related-work-comparison}
\end{table}

\textcolor{black}{\textbf{Positioning relative to prior work.} Table~\ref{tab:related-work-comparison} positions {\name} along three axes: the design stage at which PE composition is decided (fabrication vs. compilation), whether mapping is slack-aware, and whether it explicitly handles inter-iteration dependencies. Fabrication-time approaches enrich the PE itself. 
This design-time specialization lineage traces back to configurable computation accelerators (CCAs) for instruction-set customization. Clark et al.~\cite{clark2004,clark2005,yehia2005,clark2006,beret2011} introduce CCAs that execute selected dataflow subgraphs on a configurable array of function units, with follow-on work exploring transparent accelerator design spaces.
EGRA~\cite{egra_tvlsi, egra_sasp, ansaloni2009} brings this direction into the CGRA setting by mapping larger subgraphs onto reconfigurable multi-ALU clusters to improve compute density, while APEX~\cite{apex_asplos2023} merges frequent subgraphs into synthesized PE RTL, using timing analysis to set the pipeline depth needed to meet an aggressive frequency, effective for machine-learning and image-processing kernels with frequently occurring subgraphs. Because both lock operation boundaries at fabrication, composition is set once for the fabric and reused across kernels rather than tuned per kernel. Compilation-time approaches move this flexibility into the mapper: slack-aware scheduling~\cite{slack_aware} folds per-operation timing variation into modulo scheduling, and CGRA-Express~\cite{cgra_express} fuses operations through a bypass network. Cascade~\cite{cascade} instead pipelines interconnect paths post-PnR to raise frequency, which aids feed-forward kernels but cannot shorten loop-carried recurrences. Both exploit sub-cycle slack and are closest to {\name}, but restrict fusion to neighboring PEs, whereas {\name} forms VPEs across multiple hops by folding interconnect delay into the timing budget; They are also agnostic to loop-carried paths. 
A separate line of work targets inter-iteration dependencies directly: recurrence-aware scheduling~\cite{recurrence_aware_sigplan2009} groups operations along recurrence cycles atop edge-centric modulo scheduling (EMS) to improve schedule quality, and UE-CGRA~\cite{uecgra} accelerates loop-carried paths via PE-level DVFS but incurs significant area and power overhead (>30\%), yet neither harvests circuit-level slack. {\name} is the only compilation-time approach that is both slack-aware and explicitly recurrence-aware: it uses STA-derived delays to form virtual-PE boundaries around producer-consumer chains, particularly inter-iteration dependencies, collapsing them into single registered stages to reclaim wasted slack and resolve the inter-iteration dependencies that bound throughput.}

Our contributions include:
\begin{itemize}
    \item a methodology and detailed characterization of combinational timing paths derived from a silicon-proven CGRA to establish a high-fidelity baseline for circuit-aware mapping.
    \item \textcolor{black}{a timing-driven, recurrence-aware framework, {\name} that uses STA to form virtual-PE boundaries: it harvests sub-cycle slack and co-locates inter-iteration dependencies within fewer registered stages to improve performance and lower input-to-output latency, generating a Pareto-optimal frontier across operating frequencies.}
    \item integrating the {\name} flow into an open-source CGRA infrastructure and demonstrating its efficacy across diverse architectural configurations and real-world kernels, achieving significant performance and energy gains.
\end{itemize}
    
\textbf{Results:} Experimental results show that {\name} achieves, on average, 2.3x lower normalized cycle count, 6.3x better energy-delay product (EDP), and 2x lower input-to-output latency, while demonstrating that peak performance and efficiency can be achieved at much lower operating frequencies by composing PEs. {\name} further improves resource utilization by 2x and reduces register pressure by 2.2x compared to SOTA CGRAs.

\section{Background and Motivation}
This section first provides background on FO4 delay, before diving into a case study of a CGRA chip fabricated on a commercial 12nm node. We perform static timing analysis (STA) of the timing paths on 12nm and 40nm processes, translated into FO4 delays as a technology-independent metric, across six CGRA kernels. Our case study reveals substantial potential for composing primitives within a single cycle, motivating {\name}.

\subsection{FO4 delay: A technology-independent metric for timing}
The timing closure of modern high-performance spatial architectures depends on rigorous Static Timing Analysis (STA) to guarantee functional correctness across all process, voltage, and temperature (PVT) corners~\cite{sta_basics}. 
As illustrated in Fig.~\ref{fig:sta}, the non-linear scaling of path delays across successive technology nodes highlights the challenge of maintaining a unified timing abstraction. When delays are expressed in absolute temporal units (ps), the architectural model becomes tied to a specific process, complicating the development of hardware-aware software that must remain portable across technology generations.
Consequently, achieving high operating frequencies in advanced technology nodes requires Statistical STA (SSTA) to mitigate the effects of IR drop and crosstalk-induced jitter, which are exacerbated by the reduced noise margins of low-voltage (Vdd) operation ~\cite{ssta}.

To ensure a technology-independent evaluation of architectural merit, the Fan-out-of-4 (FO4) delay is utilized as the canonical metric for normalizing logic depth. Formally defined as the propagation delay of a CMOS inverter driving four identical inverters, FO4 encapsulates the intrinsic switching speed of a specific process~\cite{fo4}. While the absolute delay of an FO4 unit scales down --- from approximately 30ps at 65nm to sub-10ps in 7nm FinFET~\cite{fo4_numbers, fo4_horowitz} --- it serves as a normalized ``yardstick'' that remains consistent across different fabrication technologies. By expressing the critical path in terms of FO4 units ($T_{clk}/FO4$), 
the logical depth of an operator remains relatively comparable in FO4 terms across technology nodes, as illustrated in Fig.~\ref{fig:sta}. Thus, FO4 provides a stable baseline for identifying whether performance bottlenecks arise from architectural complexity or physical scaling limits, enabling effective cross-node benchmarking and hardware-aware optimizations. 
\textcolor{black}{For VPE formation, {\name} uses absolute post-layout delays for the target node. 
\textbf{FO4 has two roles: it shows the delay spread is structural, with the 12nm and 40nm series tracking within 13\% in FO4 terms (Fig.~\ref{fig:sta}), and it provides default cost models when retargeting before signed-off STA exists.}}
\subsection{Motivating case study}
\label{sec:sta_analysis}



\begin{figure*}[bt!]
    \centering
    \includegraphics[width=\textwidth]{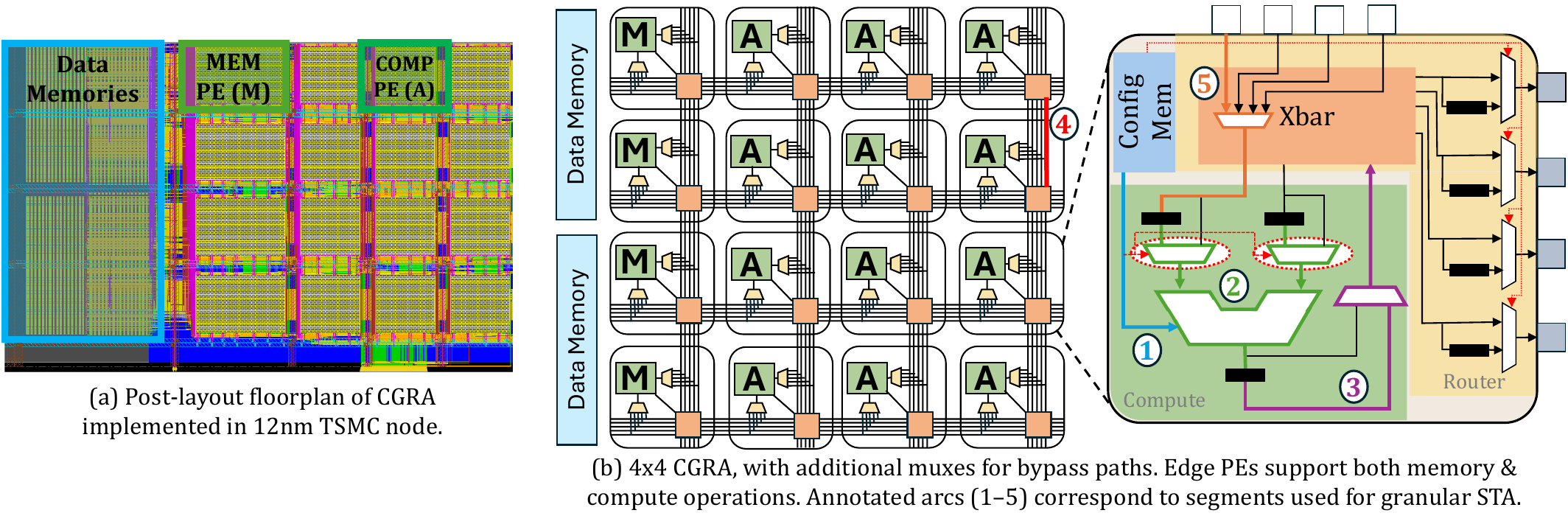}
    \caption{\textcolor{black}{Physical implementation and timing paths of a silicon-proven CGRA chip with support for {\name}.}}
    \label{fig:motivation}
\end{figure*}

Our analysis is based on a silicon-proven, energy-efficient CGRA architecture implemented in a 12nm FinFET commercial process that achieves a peak energy efficiency of 1.31 TOPS/W while meeting a target operating frequency of 1 GHz. As illustrated in Fig.~\ref{fig:motivation}(a), the floorplan shows a single $4\times4$ cluster of PEs. To provide sufficient memory bandwidth for data-intensive kernels, the four edge PEs of each cluster are specialized Memory-capable PEs (MEM) equipped with dedicated Load-Store Units (LSUs) which interface with a shared 4-port data memory. The remaining twelve internal tiles are Compute-only PEs optimized for high-throughput arithmetic kernels. This clustered, heterogeneous configuration reflects industry CGRAs like SambaNova SN10~\cite{sambanova_sn10} and Efficient Electron E1~\cite{efficient_electron_e1}, providing a representative baseline.

\textbf{Synthesis and Physical Implementation:}
The architecture was implemented and taped out in the TSMC 12nm FinFET process. Logic synthesis was performed using Cadence Genus with a multi threshold-voltage CMOS (MTCMOS) strategy. Specifically, Ultra-Low-Vt (uLVT) cells were strategically prioritized for the critical paths within the PE mesh to maximize operating frequency. The physical floorplanning and structured placement were executed in Cadence Innovus via a custom Tcl-based placement script. As illustrated in the post-layout floorplan (Fig.~\ref{fig:motivation}(a)), this script enforces strict spatial regularity by defining a 0.001$um$ coordinate grid, ensuring the fabric maintains the desired architectural symmetry. Notably, the MEM PEs are allocated a much larger silicon footprint, $\sim$1.45X the area of a standard compute PE, to accommodate the dedicated LSU logic and high-bandwidth memory port interfaces. Final routing was performed using Innovus NanoRoute, leveraging the TSMC 12nm middle metal layers between M4-M7 for low-resistance global interconnects between the four clusters, while top-metal layers M7-M9 were used for power delivery.

\textbf{Static Timing Analysis and FO4 Normalization:}
\newcommand{\circnum}[1]{\raisebox{-0.5ex}{\scalebox{1.5}{\ding{\numexpr171+#1\relax}}}}
Final timing sign-off was conducted using Cadence Tempus with post-routing parasitics (RC) input from the Standard Parasitic Exchange Format (SPEF) file. 
As illustrated in Fig.~\ref{fig:motivation}(b), the cycle time is determined by the longest PE-to-PE critical path, which we decompose into five timing segments for fine-grained characterization: 
\circnum{1} configuration memory to ALU input selection, 
\circnum{2} pure combinational ALU computation, 
\circnum{3} ALU output to the local routing crossbar, 
\circnum{4} router-to-router propagation along the programmed path, and 
\circnum{5} the final hop into the destination PE, accounting for clock skew and output registering.
Delays for the five paths were reported for TT corner with sign-off settings.

\begin{figure}[t!]
    \centering
    \includegraphics[width=\columnwidth]{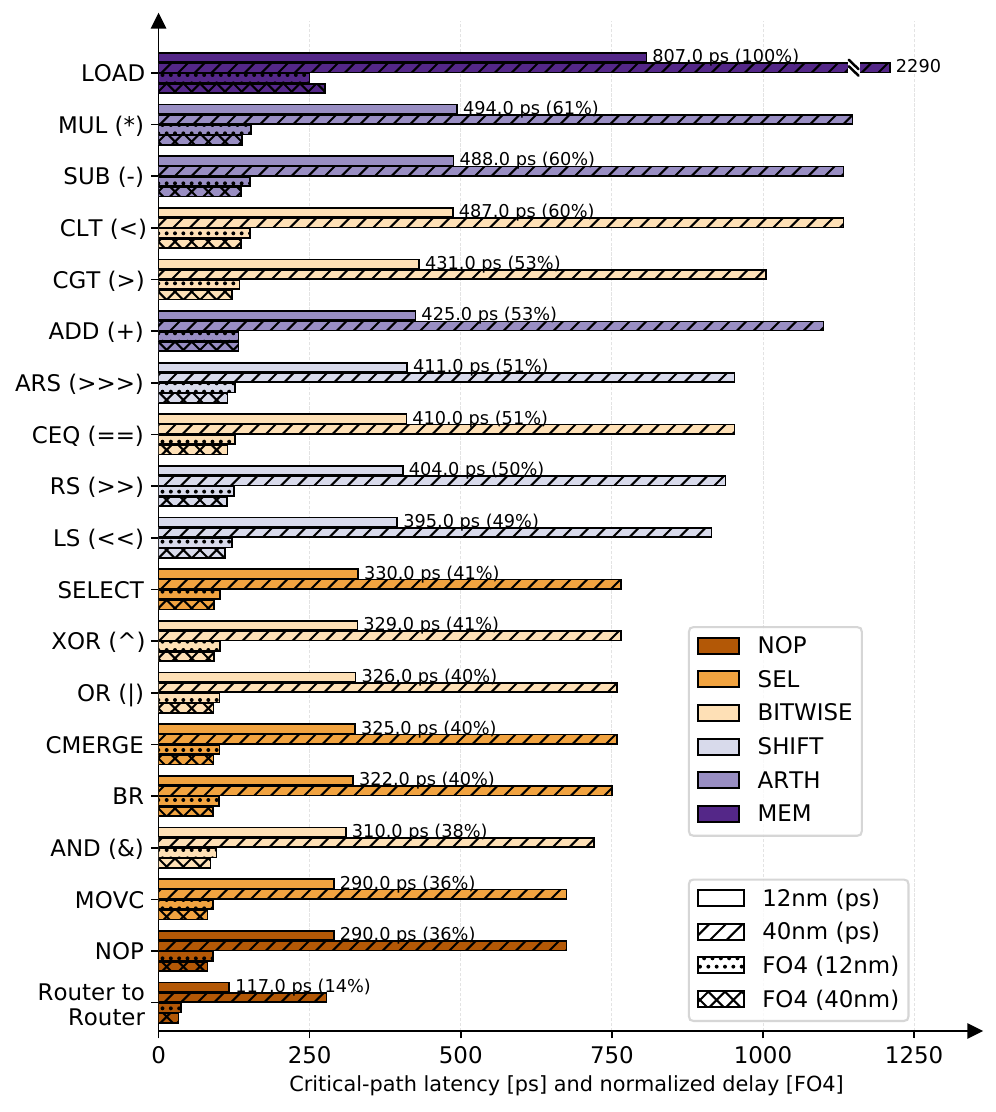}
    \caption{\textcolor{black}{Critical-path delays in 12nm and 40nm technologies. Absolute delays (ps) are shown alongside technology-normalized FO4 counts, illustrating that operator logical depth remains largely unchanged across technology nodes.}}
    \label{fig:sta}
\end{figure}

To establish a process-independent baseline, we characterize an inverter chain in the TSMC 12nm library, where each stage drives a capacitive load of four identical inverters. The resulting delay 3.24 ps defines one FO4 unit. By mapping the absolute delays from our five-arc STA across multiple PVT corners to this FO4 constant, we generate a normalized timing graph. 
To validate the consistency of this abstraction across disparate process nodes, we also performed the same physical implementation and STA flow for a 40nm UMC technology node, where one FO4 corresponds to 10.9 ps. As shown in Fig.~\ref{fig:sta}, the resulting data captures the delays for various operations and interconnect paths across 12nm, 40nm, and the corresponding FO4 metrics.

This normalization highlights the significant timing variance between different operation classes. Even within an ALU, many operations are effectively wiring or selection operations, such as \textit{MOVC}, \textit{SEXT}, \textit{SELECT}, and \textit{CMERGE} (described in Table 1), which perform no computation but are necessary for routing data across fabric. Their delay is dominated by small multiplexers and short local wires, leaving substantial slack. Next are single-level bitwise and predicate operations such as \textit{OR}, \textit{AND}, \textit{XOR}, \textit{CMP}, \textit{CGT}, and \textit{CLT}, which add one gate level and modest flag logic, yielding slightly higher but still low delay. Shifts such as \textit{RS}, \textit{ARS}, and \textit{LS} sit in the middle because a barrel shifter introduces wider mux trees and more wiring. Arithmetic primitives are the slowest by comparison: wide \textit{ADD} and \textit{SUB} are limited by carry propagation, while \textit{MUL} is typically the longest path and often sets the ALU's critical-path. Memory operations stand apart as the most significant bottleneck; due to the cumulative delay of memory macros, arbitration logic, and LSU overhead, these operations are so long they typically require two clock cycles to complete. Fig.~\ref{fig:sta} shows the variation in critical-path-delay for all these paths. \textbf{This spread explains why a uniform cycle budget wastes slack on most nodes and motivates composing fast primitives within a single cycle.}

\subsection{Analysis of applications}
\begin{figure*}[t!]
    \centering
    \includegraphics[width=\textwidth]{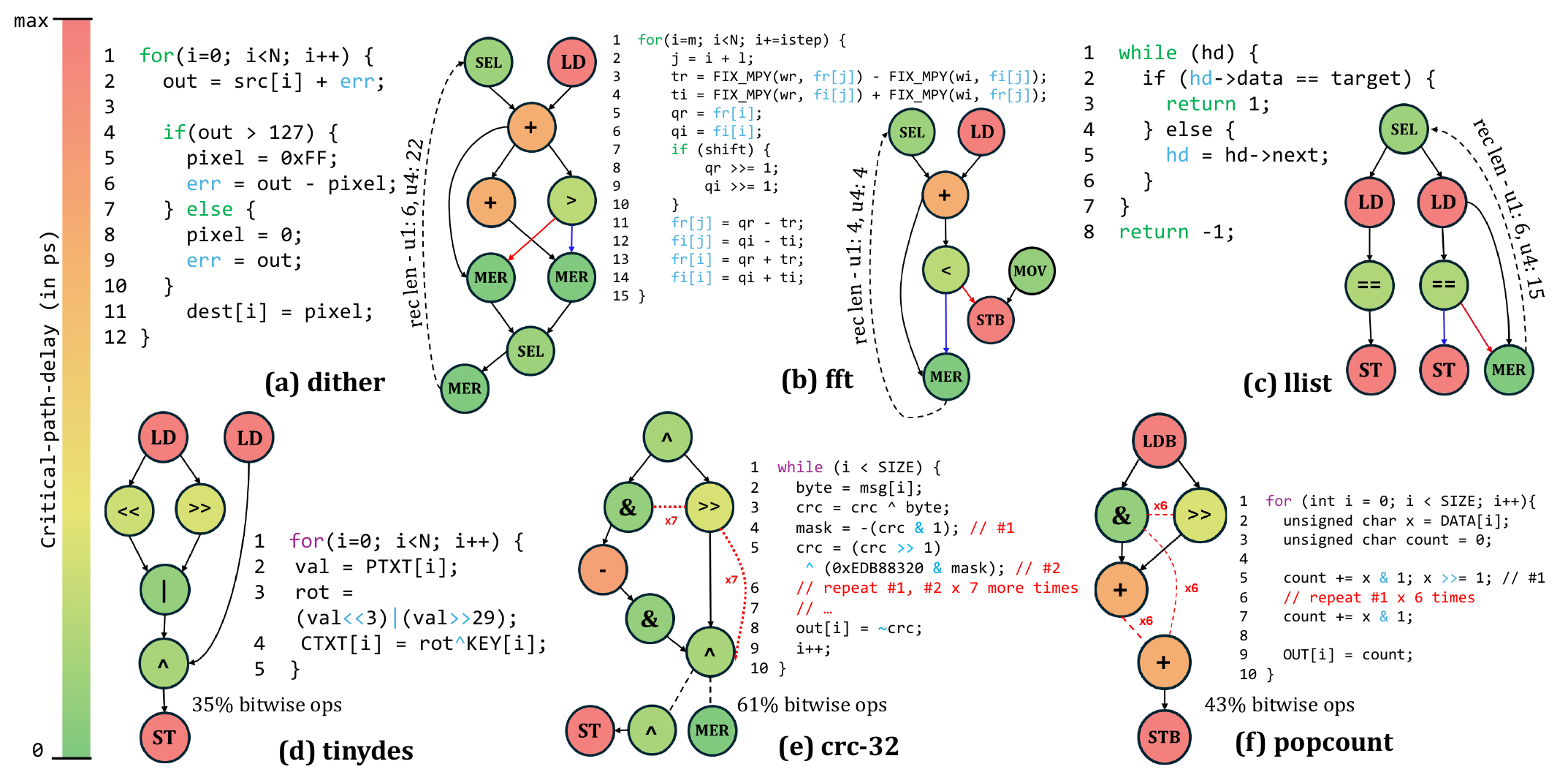}
    \caption{Six kernels with code and dataflow graphs. Nodes are colored by per-node critical-path-delay (scale at left: {\color{green}green} low, {\color{red}red} high); the dashed edge marks the loop-carried dependence. \textit{Dither}, \textit{fft}, and \textit{llist} show long back edges (recurrence lengths written for unroll 1 and 4) that bind the initiation interval, whereas \textit{tinydes}, \textit{crc-32}, and \textit{popcount} are dominated by bitwise and select nodes, leaving substantial slack.}
    \label{fig:application_analysis}
\end{figure*}

Fig.~\ref{fig:application_analysis} characterizes six kernels with their source code and dataflow graphs. Each node is colored by its intrinsic critical-path delay based on the aforementioned post-synthesis timing data. Green nodes represent low-latency operations, while red nodes signify high-delay primitives. Dashed edges denote loop-carried dependencies, which establish the fundamental recurrence boundary. Slack is the temporal difference between the global clock period and a node’s specific execution delay. Consequently, greener nodes represent a significant underutilization of the available cycle budget.

Our analysis reveals two primary inefficiencies in standard CGRA mapping. First, in kernels with true inter-iteration data dependencies, such as the pointer-chasing in \textit{llist} or the recursive stages in \textit{fft}, the DFG contains a long loop-carried path that dictates the initiation interval ($II$). Because a new iteration can only begin every $II$ cycles, the throughput is strictly bottlenecked by the recurrence delay, regardless of the timing of individual nodes. Second, bitwise-heavy kernels such as \textit{viterbi}, \textit{crc32}, and \textit{sha256} exhibit a "slack abundance." Most nodes are green, yet the global clock frequency is penalized by a small minority of slow "red" nodes. This mismatch results in a large amount of unused timing slack in every cycle. Furthermore, the standard practice of writing every intermediate result to a PE-boundary register before the next consumer executes introduces mandatory register file energy overheads that do not contribute to functional throughput. Regular kernels such as \textit{spmspm} and \textit{sddmm} also exhibit recurrence edges in their raw DFGs even though their iterations are independent. However, these are typically induction variable dependencies that can be decoupled from the core computation, modern literature and commercial CGRAs~\cite{softbrain, cascade_agu, sambanova_sn40l} mitigate this by offloading such dependencies to specialized hardware loop-counters or AGUs to maintain high throughput.

\begin{table}[bht]
\centering
\resizebox{\columnwidth}{!}{
\begin{tabular}{@{}lcccl}
\toprule
\textbf{Benchmark} & \textbf{Memory (\%)}   &   \textbf{ALU (\%)}   &   \textbf{Bitwise (\%)}   &   \textbf{Wiring (\%)}\\ \midrule
         dither & 17.9 & 17.8 & 14.3 & 50\\
         llist & 36.8 & 10.5 & 21.1 & 31.6\\
         fft & 25.4 & 29.9 & 34.3 & 10.4\\
         susan & 24.2 & 30.3 & 12.2 & 33.3\\
         bfs & 32.3	& 17.7 & 14.7	& 35.3\\
         viterbi& 23.7	& 15.8	& 15.8	& 44.7\\
         \hline
         tinydes& 26.1	& 17.4	& 34.8	& 21.7\\
         popcount& 14.3	& 28.6	& 42.9	& 14.2\\
         aes & 18.7 & 12.3 & 59 & 10 \\
         crc32& 8.2	& 18 & 60.7	& 13.1\\
         \hline
         gemm& 38.4	& 27 & 15.4	& 19.2\\
         conv2d& 25.6 & 43.6 & 18 & 12.8\\
         spmspm& 46.4 & 21.4 & 14.3 & 17.9\\
         sddmm& 42.9 & 28.6 & 10.7 & 17.8\\
         \bottomrule
\end{tabular}
}
\caption{Breakdown of operation types across the kernels}
\label{tab:op_mix}
\end{table}

As detailed in Table~\ref{tab:op_mix}, operation distribution exhibits high variance across application domains, demonstrating the inefficiency of a one-size-fits-all cycle budget. Cryptographic and mathematical kernels (e.g., crc32, popcount) are dominated by low-latency bitwise logic, which comprises up to 60.6\% of the DFG. Workloads bottlenecked by complex loop iteration dependencies and irregular control flow (e.g., dither, viterbi, bfs) allocate a substantial fraction of nodes to wiring and selection logic, ranging from 31\% to 50\%. Regular dense workloads (e.g., gemm, conv2d, spmspm) primarily utilize high-latency ALU (up to 43.6\%) and memory operations (up to 46.4\%). Despite this varied operation mix, memory nodes consistently dictate the global critical path~\cite{amber} even in kernels where they are numerically infrequent. This timing bottleneck originates from the cumulative delay of memory macros, memory-tile arbitration logic, and Load-Store Unit (LSU) overhead. Furthermore, in advanced process nodes, the long-range interconnect required to route data from the memory periphery across the PE array introduces high RC parasitics that frequently exceed the gate delay of the functional units. Consequently, these factors constrain the global $f_{max}$, generating significant unexploited timing slack within the highly prevalent bitwise and wiring operations mapped to the spatial fabric.

To reclaim this unexploited slack, we leverage the inherent configurability of spatial architectures. Unlike specialized accelerators where the dataflow is often hard-wired, CGRAs offers a unique opportunity for tight hardware-software co-design. As CGRAs are generalized configurable architectures, the compiler plays a central role in orchestrating execution. This requires hardware and software teams to work hand-in-hand to ensure spatial mappings remain physically viable. Such a tight coupling allows the mapper to access chip timing results that are typically opaque to software teams, effectively establishing timing as a first-class constraint within the hardware-software interface.

\textcolor{black}{Prior work has optimized the first two families well: APEX~\cite{apex_asplos2023} accelerates machine-learning and image-processing kernels by mining frequent subgraphs and synthesizing high-frequency specialized PEs, while slack-aware scheduling~\cite{slack_aware} and CGRA-Express~\cite{cgra_express} exploit the abundant slack in bitwise-heavy kernels through compile-time operation fusion. We include these workloads to demonstrate the framework's versatility across diverse operation mixes, but {\name} primarily targets kernels whose throughput is bound by inter-iteration dependencies, where neither higher frequency nor local fusion resolves the serialized loop-carried path.}
\section{Design of {\name}}
\label{sec:design}
\begin{figure}[h!]
    \centering
    \includegraphics[width=\columnwidth]{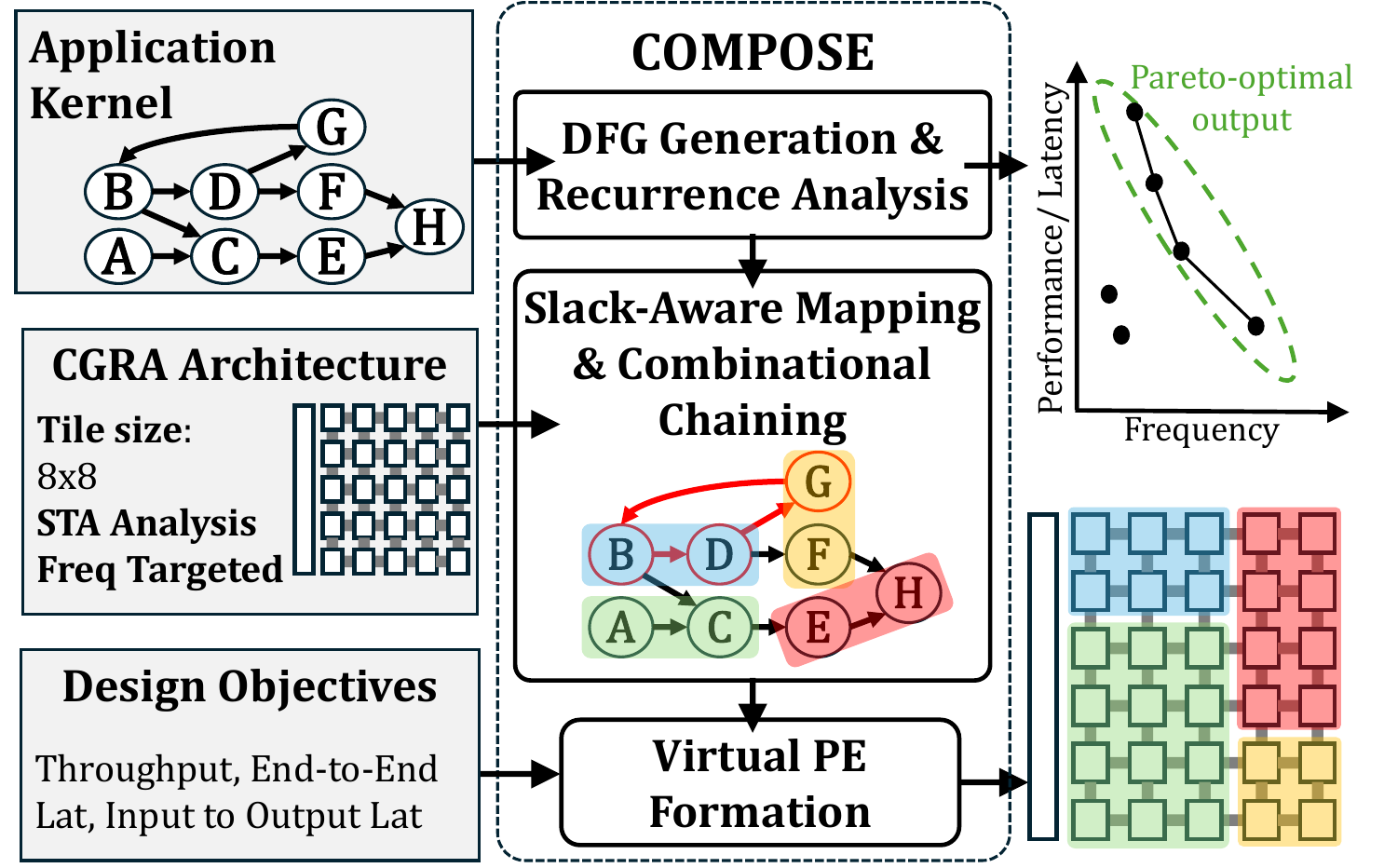}
    \caption{{\name} Framework}
    \label{fig:framework}
\end{figure}

The {\name} framework unifies circuit-level timing insights into a hardware-driven compilation strategy to optimize mapping for CGRAs. As illustrated in Fig.~\ref{fig:framework}, {\name} takes application kernels and architectural specifications, specifically post-layout STA data, as primary inputs. The framework consists of three key stages:\\
\textbf{DFG Generation \& Recurrence Analysis:} The framework initially transforms high-level kernel code into a Dataflow Graph (DFG) where nodes represent primitive operations and edges denote data-dependency. A critical sub-phase involves Recurrence Edge Analysis to identify loop-carried dependencies. These cycles fundamentally bound the Recurrence Minimum Initiation Interval (RecMII), creating a hard ceiling on steady-state throughput.\\
\textbf{Slack-Aware Mapping \& Combinational Chaining:} In this phase, the framework executes a slack harvesting strategy that integrates the total combinational delay of the data routing path with internal operation latencies. Rather than treating nodes as isolated units, the mapper evaluates the complete propagation delay, accounting for multi-hop interconnect latency and switching overheads required to reach a destination. It greedily chains producer-to-consumer operations such that their cumulative path delay, inclusive of all routing and switching overhead, is bounded by a target clock period ($T_{clk}$).\\
\textbf{Virtual PE Formation:} At compile time, {\name} defines Virtual PE (VPE) boundaries that decouple logical operation mapping from the fixed boundaries of individual PEs. By spatially fusing sub-graphs into single-cycle composites, the framework (1) reduces end-to-end execution latency, (2) improves throughput in dependency-bound kernels by collapsing multi-cycle recurrence paths into fewer cycles, and (3) minimizes data movement overhead by deferring output registration for locally consumable intermediate values, significantly reducing register file traffic to enhance energy efficiency.

The framework generates a Pareto-optimal frontier of design points, generating multiple schedules that explore the trade-offs between throughput, latency, and energy efficiency. This enables the user to select the specific schedule that best aligns with their design objectives, whether maximizing performance for compute-intensive workloads or minimizing power for energy-constrained environments.

\subsection{DFG Generation \& Recurrence Analysis}
\label{sec:dfg-gen}
\begin{figure*}[t!]
    \centering
    \includegraphics[width=\textwidth]{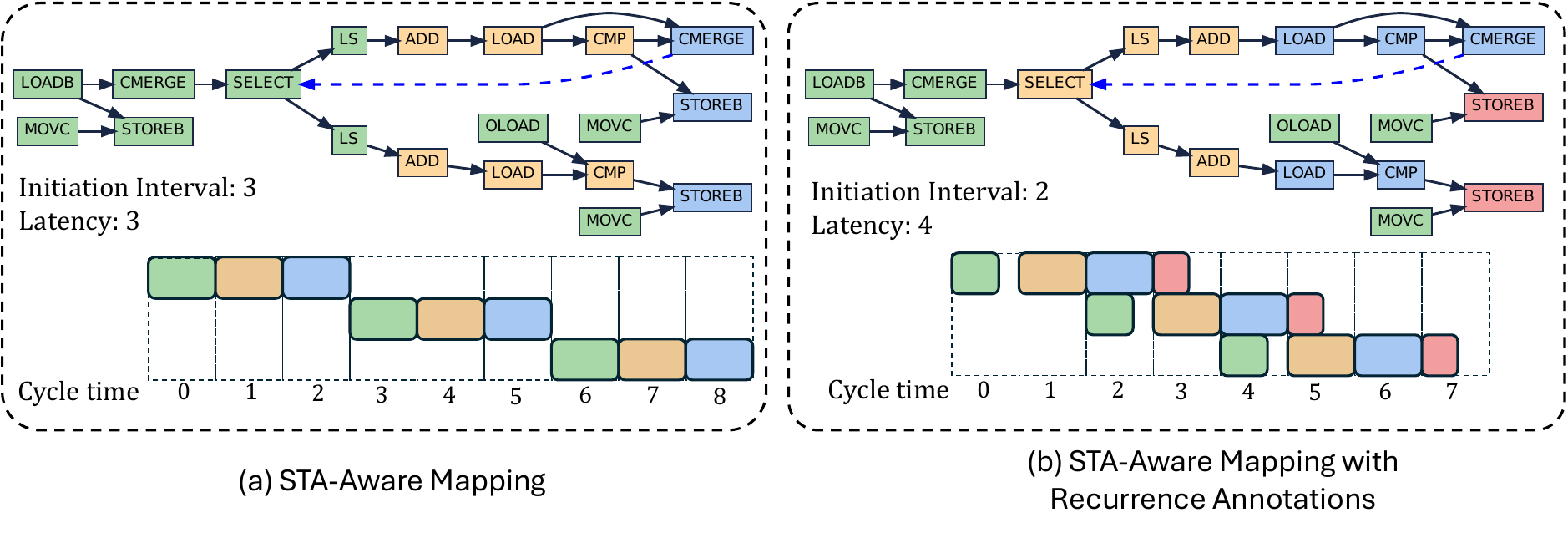}
    \vspace{-0.2cm}
    \caption{STA-Aware Mapping with and without Recurrence Co-location, where each color represents a distinct composite. (a) Without recurrence awareness, operations span 3 pipeline stages, giving II = 3. (b) Co-locating recurrence-dependent operations within a single cycle reduces II to 2, but fragments non-critical operations across an extra pipeline stage, increasing input-to-output latency from 3 to 4.}
    \label{fig:recurrence-tradeoff}
\end{figure*}

\begin{algorithm}[t]
\linespread{1.0}
\DontPrintSemicolon
\small
\SetAlgoSkip{}
\SetAlgoVlined
\SetCommentSty{textrm}
\newcommand{\mycomment}[1]{\textcolor{red}{$\triangleright$ #1}}
\small

\caption{DFG Generation and Recurrence Analysis}
\label{alg:dfg-gen}

\KwIn{Loop body $L$, Instruction set $I$, Control Flow Graph $\textit{CFG}$}
\KwOut{DFG $G(V, E)$, Recurrence map $\textit{RecII}$}

$V \leftarrow \emptyset$; \quad $E \leftarrow \emptyset$; \quad $\textit{RecII} \leftarrow \emptyset$\;

\mycomment{Step 1: Identify CFG back-edges \& build forward-reachability sets}\;
$\textit{BackEdges} \leftarrow \text{FindBackEdges}(\textit{CFG})$\;
\For{each basic block $B$ in $\textit{CFG}$}{
    $\textit{FwdReach}[B] \leftarrow \text{DFS}(B, \textit{CFG} \setminus \textit{BackEdges})$ \mycomment{BBs reachable without crossing back-edges}\;
}

\mycomment{Step 2: Create DFG nodes with post-layout timing}\;
\For{each instruction $i \in I$}{
    $v \leftarrow \text{CreateNode}(i)$\;
    $\text{Annotate}(v, \text{STA}(i))$ \mycomment{attach post-layout delay $\delta(v)$}\;
    $V \leftarrow V \cup \{v\}$\;
}

\mycomment{Step 3: Edge construction and recurrence identification}\;
\For{each $v \in V$}{
    \For{each operand of $v$ produced by $u \in V$}{
        $E \leftarrow E \cup \{(u, v)\}$\;
        \uIf{$\text{BB}(v) \notin \textit{FwdReach}[\text{BB}(u)]$}{
            $\textit{RecII}[(u,v)] \leftarrow 1$ \mycomment{loop-carried: destination not forward-reachable from source}\;
        }
        \Else{
            $\textit{RecII}[(u,v)] \leftarrow 0$ \mycomment{intra-iteration edge}\;
        }
    }
}
\Return{$G(V, E),\, \textit{RecII}$}\;
\end{algorithm}

Loop-carried dependencies (recurrences) impose a fundamental throughput bottleneck on CGRA execution: if a recurrence path spans $d$ pipeline stages, the architecture cannot initiate a new iteration until $d$ cycles have elapsed, setting $\mathit{RecMII} \geq d$~\cite{modulo_scheduling}. By identifying recurrence edges at the DFG level and co-locating their endpoints within a single VPE, {\name} collapses multi-stage recurrence paths into few registered cycles, directly reducing $\mathit{RecMII}$ and improving steady-state throughput.

However, this recurrence-first strategy introduces a latency tradeoff. As illustrated in Fig.~\ref{fig:recurrence-tradeoff}, prioritizing recurrence co-location can force a new VPE to be instantiated at Node~2 despite underutilized slack in VPE~0, solely to ensure the recurrence completes within a single cycle wherever possible. While this maximizes throughput for dependency-bound loops, it fragments non-critical operations across additional pipeline stages, increasing end-to-end latency. {\name} exposes this tradeoff by generating multiple mapping points across the throughput-latency Pareto frontier, allowing the user to select the schedule that best matches their design objective.

The first stage of the {\name} compiler transforms a loop body into a DFG annotated with post-layout timing and recurrence information, as presented in Algorithm~\ref{alg:dfg-gen}. The algorithm begins by identifying all control-flow back-edges in the loop's CFG using a standard depth-first traversal (Step~1). For each basic block $BB$, a forward-reachability set $\textit{FwdReach}[BB]$ is constructed by performing a DFS over the CFG while excluding back-edges. This set captures all basic blocks reachable from $BB$ within a single loop iteration, and serves as the basis for distinguishing intra-iteration edges from loop-carried (recurrence) edges.

Next, each instruction is instantiated as a DFG node and annotated with its post-layout operation latency $\delta(v)$, obtained from gate-level STA (Step~2). During edge construction (Step~3), the algorithm classifies each data dependence: if the destination basic block $\text{BB}(v)$ is not in $\textit{FwdReach}[\text{BB}(u)]$, the edge $(u,v)$ is loop-carried and marked $\textit{RecII}[(u,v)] = 1$; otherwise it is an intra-iteration edge with $\textit{RecII}[(u,v)] = 0$. This forward-reachability test cleanly handles both explicit PHI-node recurrences and implicit topological back-edges without requiring pattern-specific detection logic.

The recurrence map $\textit{RecII}$ directly informs the downstream mapper in two ways. First, recurrence edges define hard constraints on the initiation interval: operations connected by recurrence edges must complete within a bounded number of cycles, establishing the $\mathit{RecMII}$ lower bound on $\mathit{II}$. Second, the mapper uses recurrence edges to co-locate producer-consumer pairs into the same Virtual PE, ensuring that loop-carried paths complete within a single registered stage wherever possible.

A fundamental challenge is determining whether to partition the DFG into single-cycle composites prior to or during the physical mapping phase. While ahead-of-time partitioning based on critical-path delay might simplify the mapper flow, it fails to account for the stochastic nature of spatial resource contention. Specifically, partitioning at the DFG level cannot guarantee hardware feasibility within a single cycle because physical resource constraints---primarily routing congestion---are not visible until the mapping is done. This approach is essential for architectures with single-cycle, multi-hop interconnects, where routing overhead is highly variable~\cite{peh_noc} and sensitive to the specific spatial placement of PEs.

To address this, {\name} defers partitioning to the mapping stage. If the mapper encounters a resource bottleneck that prevents a node from being placed within the current timing-valid composite, it automatically registers the output and initiates a new VPE. By deferring composition until physical placement, the framework accounts for actual routing delays---including multi-hop interconnect latency $d_{\mathrm{hop}}$ from post-layout PnR data---ensuring single-cycle chains are practically realizable within the global $T_{\mathrm{clk}}$ constraint.

\subsection{Slack-Aware Mapping \& Combinational Chaining}
\begin{algorithm*}[t]
\linespread{0.95}
\DontPrintSemicolon
\SetAlgoSkip{}
\small
\SetAlgoVlined
\SetCommentSty{textrm}
\newcommand{\mycomment}[1]{\textcolor{red}{$\triangleright$ #1}}

\caption{Slack-Aware Virtual PE Mapping}
\label{alg:combi-map}

\KwIn{$G(V, E)$, $\textit{RecII}$: Annotated DFG from Alg.~\ref{alg:dfg-gen}. $\textit{RecII}[(u,v)]\!\in\!\{0,1\}$: 1 for recurrence, 0 for intra-iteration. $\mathcal{A}$: CGRA ($X\!\times\!Y$). $\delta$: Post-layout op.\ latencies (ps). $d_{\mathrm{hop}}$: Per-hop interconnect delay (ps, from PnR). $T_{\mathrm{clk}}$: Clock period (ps).}
\KwOut{$\mathcal{M}$: Node-to-PE mapping. $\mathit{II}$: Initiation interval. $L$: Pipeline latency (number of VPE stages).}

\BlankLine
\mycomment{Phase 1: STA and recurrence analysis --- establish mapping order and recurrence groups}\;
\For{each $v \in V$ in topological order over forward edges ($\textit{RecII}[(u,v)]\!=\!0$)}{
    $\textit{arr}[v] \leftarrow \delta(v) + \max_{(u,v)\in E,\,\textit{RecII}[(u,v)]=0} \textit{arr}[u]$ \mycomment{cumulative delay from sources; recurrence edges skipped}\;
}
Initialize Union-Find $\mathcal{U}$ over $V$\;
\lFor{each $(u,v) \in E$ where $\textit{RecII}[(u,v)]\!=\!1$}{$\mathcal{U}.\text{Unite}(u,v)$ \mycomment{group recurrence-connected nodes}}
Sort $V$ by $\textit{arr}[v]$ in ASAP order\;

\BlankLine
\mycomment{Phase 2: Compute initial II from recurrence structure}\;
$\textit{RecMII} \leftarrow 0$\;
\lFor{each recurrence group $C \in \mathcal{U}$}{$\textit{RecMII} \leftarrow \max(\textit{RecMII},\; \lceil \sum_{v \in C} \delta(v) / T_{\mathrm{clk}} \rceil)$ \mycomment{min cycles for recurrence group}}
$\mathit{II} \leftarrow \max(\textit{RecMII}, 1)$\;

\BlankLine
\mycomment{Phase 3: Incremental VPE formation during mapping}\;
\While{$\mathit{II} \leq \mathit{II}_{\max}$}{
    \For{$r = 1$ \KwTo $R_{\max}$}{
        Instantiate $\mathcal{A}_{\mathit{II}}$: $T_{\max}\!=\!\mathit{II}$, $\textit{minLatBetweenPEs}\!=\!0$, $\textit{maxHops}\!\geq\!X\!+\!Y$\;
        $k \leftarrow 0$;\; $\textit{delay}[0] \leftarrow 0$;\; $\textit{success} \leftarrow \textbf{true}$\;
        \For{each node $v \in V$ in ASAP order}{
            \lIf{$\exists\, u \in \mathcal{U}.\text{Group}(v)$ already in VPE $k'\!\neq\!k$}{$\textit{success} \leftarrow \textbf{false}$; \textbf{break} \mycomment{recurrence group split: escalate II}}
            $(\textit{placed}, \textit{hops}) \leftarrow \text{PlaceAndRoute}(v, \mathcal{A}_{\mathit{II}},\, k \bmod \mathit{II})$ \mycomment{attempt placement; returns status and hop count}\;
            $\textit{pathDelay} \leftarrow \textit{delay}[k] + \textit{hops} \times d_{\mathrm{hop}} + \delta(v)$ \mycomment{total combinational path: accumulated + routing + op}\;
            \If{not placed $\,\mathbf{or}\,$ $\textit{pathDelay} > T_{\mathrm{clk}}$}{
                \lIf{placed}{Undo placement of $v$ \mycomment{routable but timing violated by routing delay}}
                $k \leftarrow k + 1$;\; $\textit{delay}[k] \leftarrow 0$ \mycomment{register output of current VPE, start new VPE}\;
                $(\textit{placed}, \textit{hops}) \leftarrow \text{PlaceAndRoute}(v, \mathcal{A}_{\mathit{II}},\, k \bmod \mathit{II})$\;
                \lIf{not placed}{$\textit{success} \leftarrow \textbf{false}$; \textbf{break}}
                $\textit{pathDelay} \leftarrow \textit{hops} \times d_{\mathrm{hop}} + \delta(v)$\;
            }
            $\textit{vpe}[v] \leftarrow k$;\; $\textit{delay}[k] \leftarrow \textit{pathDelay}$\;
        }
        \lIf{success}{$L \leftarrow k+1$; \Return{$\mathcal{M}, \mathit{II}, L$}}
    }
    $\mathit{II} \leftarrow \mathit{II} + 1$ \mycomment{escalate II: more time-slots to accommodate VPEs}\;
}
\Return{Failure}\;
\end{algorithm*}

Algorithm~\ref{alg:combi-map} outlines the mapper algorithm, which begins by performing a forward STA over the acyclic subgraph of the DFG---all edges excluding recurrence edges identified in Section~\ref{sec:dfg-gen} (Phase~1 of Alg.~\ref{alg:combi-map}). For each node $v$, the cumulative arrival time $\textit{arr}[v]$ is computed as the maximum arrival among its forward predecessors plus its own post-layout operation latency $\delta(v)$, sourced from gate-level STA data. Recurrence edges are skipped in this traversal since their values arrive from the previous loop iteration via a latch.

Simultaneously, the mapper establishes recurrence groups using a Union-Find structure: all nodes connected by recurrence edges are unified into equivalence classes. Nodes are then sorted by arrival time in ASAP order to establish the mapping priority. Phase~2 computes the initial $\mathit{II}$ by evaluating each recurrence group $C$: the minimum number of clock cycles required to execute all operations in $C$ is $\lceil \sum_{v \in C} \delta(v) / T_{\mathrm{clk}} \rceil$, and $\mathit{RecMII}$ is the maximum across all groups. This provides a tight lower bound on the initiation interval before any physical mapping is attempted.
\vspace{-0.1cm}
\subsection{Virtual PE Formation}
Rather than partitioning the DFG into VPEs ahead of time, {\name} forms VPEs incrementally during the physical mapping stage (Phase~3 of Alg.~\ref{alg:combi-map}). The mapper processes nodes in ASAP order and greedily extends the current VPE $k$ by attempting to place each node onto the CGRA at time-slot $k \bmod \mathit{II}$. After placement, the total combinational path delay is evaluated as $\textit{delay}[k] + \textit{hops} \times d_{\mathrm{hop}} + \delta(v)$, where $d_{\mathrm{hop}}$ is the per-hop interconnect delay obtained from post-layout PnR data, and \textit{hops} is the physical routing distance returned by the placer. If the accumulated path delay remains within $T_{\mathrm{clk}}$, the node is committed to the current VPE. If the delay exceeds $T_{\mathrm{clk}}$, or if placement fails due to routing congestion, the mapper registers the output of the current VPE, increments $k$, and initiates a new VPE---effectively inserting a pipeline register boundary. The node is then re-placed in the new VPE. A critical constraint enforces recurrence correctness: if a node belongs to a recurrence group that has already been partially mapped to a different VPE, the mapper immediately declares the current $\mathit{II}$ infeasible and escalates, since splitting a recurrence group across VPEs would violate the single-cycle recurrence requirement.

Upon successful placement of all nodes, the pipeline latency is $L = k + 1$ stages. If mapping fails at the current $\mathit{II}$, the framework increments $\mathit{II}$, which provides additional time-slots and PE resources, and retries. This produces a set of valid mapping points across increasing $\mathit{II}$ values, enabling exploration of the throughput-latency Pareto frontier.

\textcolor{black}{\subsection{Microarchitectural support for {\name}}
\begin{figure}[t!]
    \centering
    \includegraphics[width=\columnwidth]{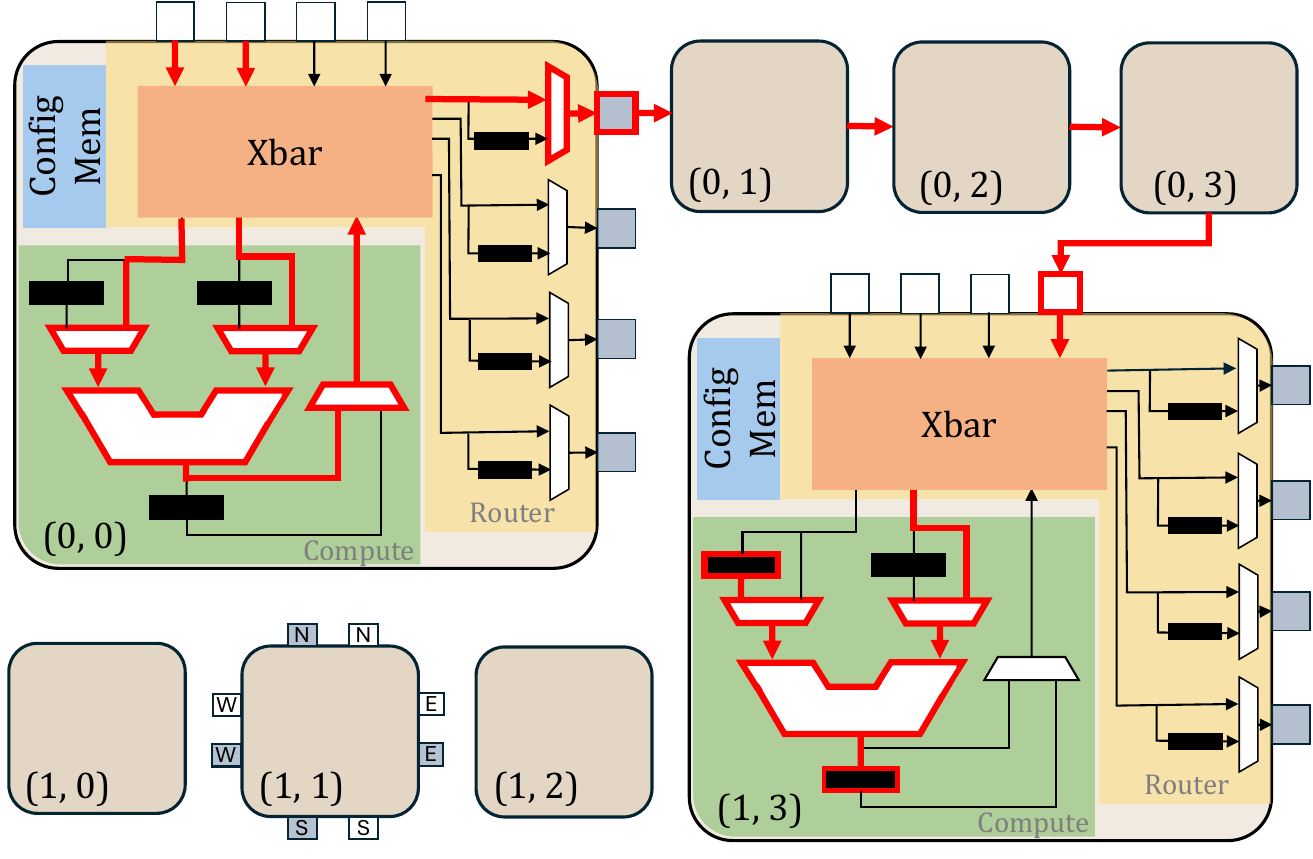}
    \caption{\textcolor{black}{Spatial composition of 2 PEs across 4 hops into 1 VPE. The interconnect microarchitecture enables combinational traversal through intermediate PEs, while ALU input ($I1$,$I2$) and output ($RES$) bypass multiplexers facilitate the chaining of multiple operations. Standard registers, indicated in black, are bypassed to ensure single cycle propagation.}}
    \label{fig:microarch}
\end{figure}
To fully realize the dynamic construction of VPEs generated by the {\name} framework, microarchitectural support is required at both the interconnect and PE datapath levels. The framework is parametrically adaptable to various NoC routing microarchitectures, accommodating fabrics that employ either standard single-cycle single-hop or single-cycle multi-hop routing. 
Because these flexible routing capabilities are already prevalent in many established academic and commercial CGRAs~\cite{pace_isocc, efficient_electron_e1, amber, snafu, riptide}, {\name} is broadly applicable across existing chips. However, while general CGRAs support multi-hop routing, they do not natively support unlatched combinational chaining. 
Therefore, explicit modifications must be added to the PE datapath. Specifically, executing dependent producer-consumer pairs within the same clock cycle requires the addition of two bypass multiplexers at the input stage of each ALU. 
These multiplexers enable the functional unit to dynamically select between the latched data from the localized register file and the direct, unlatched operand forwarded combinationally from an upstream producer (as shown in Fig.~\ref{fig:microarch}). For the 4$\times$4 Generic CGRA characterized in Section 2, integrating these input bypass multiplexers incurs a marginal hardware footprint, resulting in an area overhead of only 3.8\% and a static power increment of 2.3\% (see Section 5.4 for details). 
Crucially, this negligible structural overhead is substantially offset by significant reductions in overall dynamic power, as combinational chaining inherently eliminates the switching activity associated with intermediate register file read and write operations across the execution path.}
\section{Evaluation Methodology}
We describe the experimental setup and methodology to evaluate the end-to-end framework for {\name} in this section. Our STA analysis toolflow was previously detailed in Section~\ref{sec:sta_analysis}.


\subsection{Compilation and Mapping}
{\name} uses the Morpher toolchain~\cite{agile} which includes dataflow graph generation, mapping, and a cycle accurate simulator written in C++. {\name} operates in the mapping stage of the Morpher toolchain. It generates static configurations for a given target architecture, clock frequency, and operation timings. 
\textcolor{black}{The per-operation and per-hop delays are post-PnR STA values with SPEF parasitics from the taped-out chip, signed off at the Slow-Slow corner with a 5\% margin. Per-hop delay does not accumulate with hop count, as each intermediate bypass PE re-drives the signal to full logic level through its crossbar. We verified this on a chain of 8 PEs and observed no degradation across hops.}
The mapping process of the kernels typically takes a few minutes based on the size of the DFG. Since scheduling is static, the performance is deterministic and known at compile time. {\name}'s mapping depends on the targeted cycle time, so, we evaluate $T_{clk}$ from 100MHz to 1GHz. 
This range is representative of existing edge CGRA systems that operate at relatively modest frequencies, e.g. SNAFU at $\sim$50MHz~\cite{snafu} and Amber~\cite{amber} operates at $\sim$500MHz. These lower clock frequencies suggest the presence of mappings with non-trivial timing slack which our approach seeks to exploit.
To assess scalability, we also evaluate the framework on a larger 8$\times$8 CGRA.

\subsection{Baselines}
We evaluate {\name} against a Generic CGRA baseline and two intermediate variants, each refining the previous approach. 
\textcolor{black}{The \textbf{Generic CGRA}, modeled after HyCube~\cite{agile, pace}, uses simulated annealing (SA) based modulo scheduling~\cite{modulo_scheduling}, where one instruction executes per PE per cycle.} 
\textcolor{black}{We also include a CGRA-Express-like baseline that introduces compile-time operation fusion via a bypass network, following the approach of CGRA Express~\cite{cgra_express}. Fusion is restricted to neighboring PEs only and does not account for recurrence dependencies.}
\textbf{{\name} (Pre-Map)} introduces time-driven decomposition by partitioning the DFG based on the target clock period and mapping each partition independently with SA. Since partitioning precedes mapping, feasibility is not guaranteed, leading to fragmentation and poor utilization. \textbf{{\name} (In-Map)} interleaves partitioning and mapping greedily, improving flexibility, but does not account for recurrence dependencies or directly optimize II. Finally, {\name} extends In-Map with dynamic, recurrence-aware partitioning and mapping, incorporating loop-carried dependencies to more effectively reduce II by prioritizing full loop-carried paths within a cycle where feasible.

\subsection{Kernels Used}
\begin{table}[ht!]
\centering
\resizebox{\columnwidth}{!}{
\setlength{\tabcolsep}{3pt} 
\begin{tabular}{@{} p{1.6cm} l l cc cc @{}}
\toprule
\multirow{2}{*}{\textbf{Category}} & \multirow{2}{*}{\textbf{Kernel}} & \multirow{2}{*}{\textbf{Description}} & \multicolumn{2}{c}{\textbf{No. of nodes}} & \multicolumn{2}{c}{\textbf{Recur. length}} \\ \cmidrule(lr){4-5} \cmidrule(lr){6-7}
 &  &  & \textbf{u1} & \textbf{u4} & \textbf{u1} & \textbf{u4} \\ \midrule
\multirow{6}{1.6cm}{Loop-carried path} 
                                   & dither   & image dithering algo.     & 28 & 64 & 6 & 22\\
                                   & llist    & linked-list search        & 19 & 55 & 6 & 15\\
                                   & fft      & fast fourier transform    & 67 & 227 & 4 & 4\\
                                   & susan    & image smoothing algo.     & 33 & 78 & 4 & 6\\
                                   & bfs      & graph bfs                 & 34 & 136 & 6 & 18\\
                                   & viterbi  & viterbi decoding          & 38 & 76 & 4 & 4\\
                                    \cmidrule(l){1-7}
\multirow{5}{1.6cm}{Bitwise-heavy}     
                                   & tinydes  & des encryption            & 23 & 52 & 4 & 3\\
                                   & popcount & population count          & 35 & 113 & 4 & 3\\
                                   & crc32    & 32-bit crc                & 61 & 211 & 24 & 90\\
                                   & aes    & aes encryption                & 171 & 591 & 10 & 42\\ \cmidrule(l){1-7}
\multirow{4}{1.6cm}{Linear algebra \& AI}    
                                   & gemm     & dense matrix mult.        & 26 & 60 & 4 & 3\\
                                   & conv2d   & 2d convolution            & 39 & 91 & 4 & 3\\
                                   & spmspm   & sparse-sparse mult.       & 28 & 71 & 4 & 4\\
                                   & sddmm    & sampled dense mult.       & 28 & 71 & 4 & 5\\ \bottomrule
\end{tabular}
}
\caption{Kernels used for evaluation and their descriptions.}
\label{tab:kernels}
\end{table}
{\name} is evaluated across 14 kernels, categorized into three distinct classes to demonstrate its versatility. (1) Kernels with important inter-iteration dependencies include dither, llist, fft, susan, bfs, and viterbi. These applications possess \textbf{long loop-carried paths} that typically dictate the initiation interval, providing a rigorous test for the efficiency of the underlying mapping framework. (2) \textbf{Bitwise-heavy} kernels comprise of tinydes, popcount, aes, and crc32. Because bitwise operations have low latency, they often result in significant slack wastage. (3) \textbf{Linear Algebra and AI} kernels such as gemm, conv2d, spmspm, and sddmm represent the highly regular, compute-intensive workloads with independent loops prevalent in modern AI workloads. The kernels and their characteristics are summarized in Table~\ref{tab:kernels}.

\section{Evaluation Results}
\begin{figure*}[t!]
	\centering
	\includegraphics[width=\textwidth]{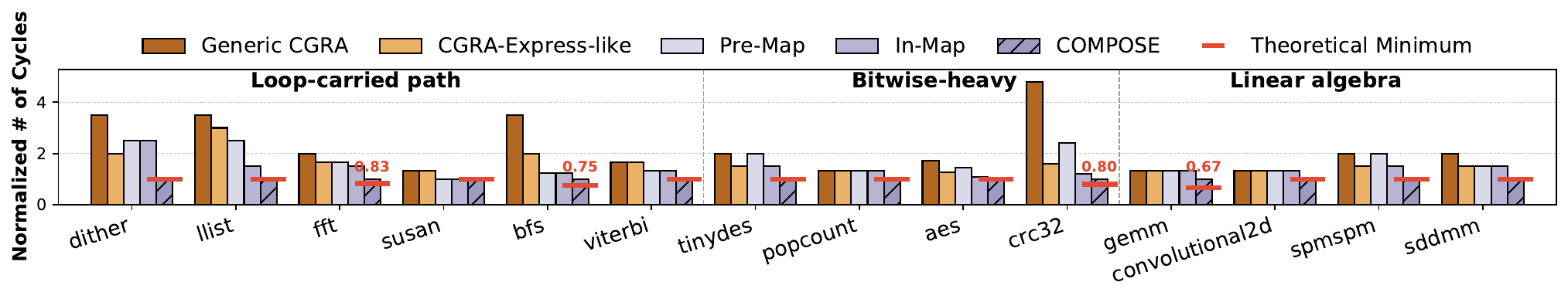}
    \vspace{-0.2cm}
	\caption{\textcolor{black}{Normalized cycle count across workloads comparing baselines and {\name} against the theoretical minimum.}}
	\label{fig:perf}
\end{figure*}

\begin{figure*}[t!]
	\centering
	\includegraphics[width=\textwidth]{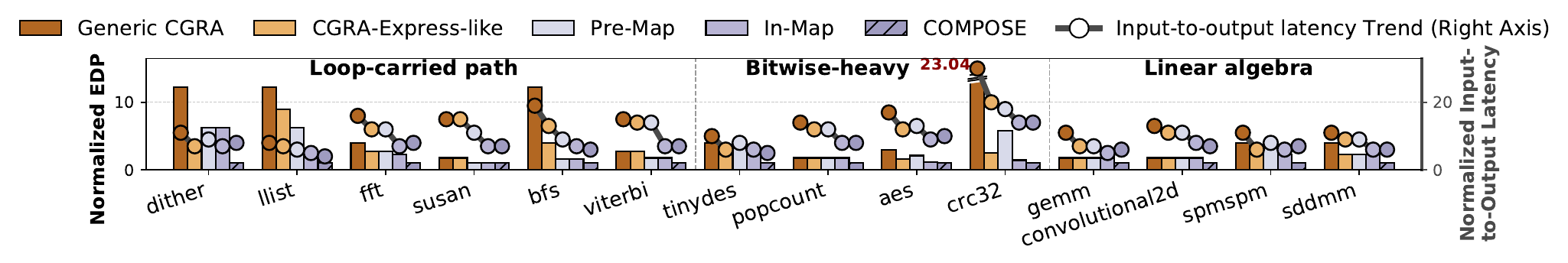}
    \vspace{-0.5cm}
	\caption{\textcolor{black}{Normalized EDP and input-to-output latency across workloads for all baselines and {\name}.}}
	\label{fig:edp}
\end{figure*}


\subsection{{\name} achieves near-optimal performance \& energy efficiency}
\textbf{Performance:} Fig.~\ref{fig:perf} compares normalized cycle counts across Generic CGRA, Pre-Map, In-Map, and {\name} against the minimum achievable bound. 
{\name} consistently approaches the theoretical lower bound, outperforming the baseline and variants by integrating timing-aware composition with recurrence-aware mapping. On average, {\name} reduces cycles by 2.3x over Generic CGRA, \textcolor{black}{1.6x over CGRA-Express-like}, 1.7x over Pre-Map and 1.4x over In-Map, with gains mostly pronounced in workloads where either recurrence constraints or slack underutilization dominate. 
\textcolor{black}{Against the theoretical minimum $(nodes/PE_{count})$, which no schedule can beat, {\name} lands within 6.8\% on average.}

\begin{figure}[h]
	\centering
	\includegraphics[width=\columnwidth]{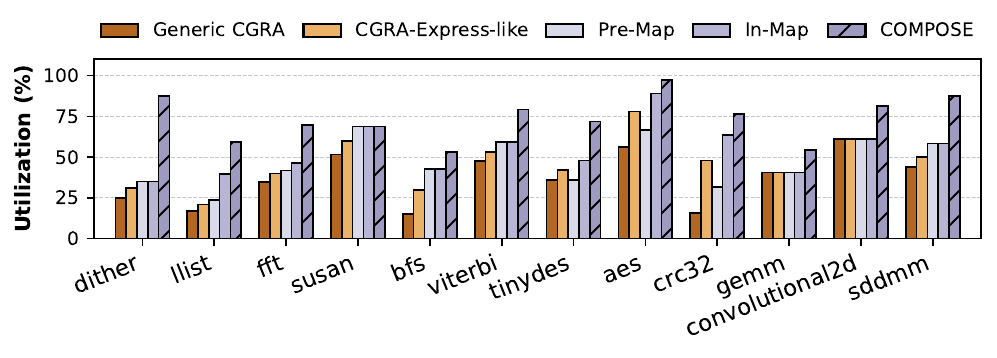}
    \vspace{-0.6cm}
	\caption{\textcolor{black}{Improved PE utilization with {\name}.}}
	\label{fig:utilization}
\end{figure}

Pre-Map introduces timing-driven partitioning but often suffers from infeasible partitions that must be split during mapping, leading to fragmentation and sometimes even degraded cycle counts. In-Map improves feasibility by composing during mapping, but without recurrence awareness, it fails to optimally reduce initiation intervals. {\name} unifies both aspects, enabling physically realizable compositions while explicitly co-locating recurrence paths. 
\textcolor{black}{This is further reflected in higher PE utilization (Fig.~\ref{fig:utilization}), where {\name} consistently achieves better resource utilization than all baselines,
including CGRA-Express-like baseline, since longer cross-hop chains complete more operations per active cycle.}

The impact is most visible in loop-carried workloads such as bfs and dither, where {\name} reduces cycle count by collapsing recurrence paths into fewer pipeline stages. In contrast, Generic CGRA is limited by recurrence-induced serialization, executing dependent operations across multiple cycles despite available spatial resources. 
\textcolor{black}{CGRA-Express-like baseline keeps input-to-output latency low because its fusion is inherently short, but on recurrence-bound kernels like dither, these short chains cannot collapse the loop-carried path, so it sacrifices the initiation interval and throughput that {\name} achieves.} 
Pre-Map and In-Map also fall short due to their inability to consistently co-locate recurrence-dependent operations, leaving a gap to the minimum bound. In bitwise-heavy workloads such as crc32, {\name} achieves significant gains by exploiting slack and packing multiple low-latency operations within a single cycle, while Pre-Map and In-Map only partially utilize this due to fragmentation or lack of global timing awareness. In contrast, workloads like gemm show minimal improvement across all methods, including {\name}, as execution is constrained by available parallelism and PE resources. \textbf{{\name} improves performance by reducing cycle count through timing-aware composition and explicit handling of recurrence dependencies.}

\textbf{Energy Efficiency:} Fig.~\ref{fig:edp} shows the normalized Energy-Delay Product (EDP) across all variants. 
{\name} consistently achieves the lowest EDP, with improvements of 6.3x over Generic CGRA, \textcolor{black}{2.9x over CGRA-Express-like}, 3x over Pre-Map and 2.1x over In-Map. 
\textcolor{black}{CGRA-Express-like baseline registers most intermediate values because its chains terminate early, and these extra register writes consume the switching energy that {\name} avoids by forwarding values within longer composed chains.}
While Pre-Map and In-Map reduce delay, their energy savings are limited due to additional register activity introduced because of poor partitioning and mapping. {\name}, in contrast, reduces both delay and switching activity, leading to more consistent EDP improvements across workloads.

\begin{figure}[h!]
	\centering
	\includegraphics[width=\columnwidth]{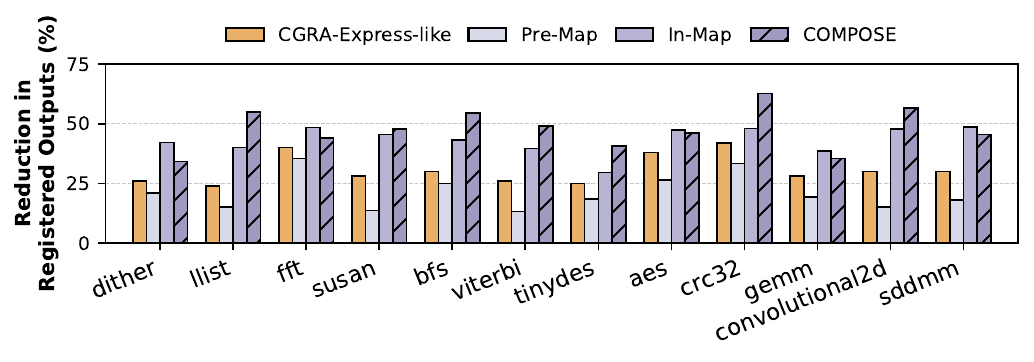}
    \vspace{-0.6cm}
	\caption{\textcolor{black}{Reduction in intermediate register writes compared to Generic CGRA.}}
	\label{fig:register}
\end{figure}
Fig.~\ref{fig:register} shows the reduction in intermediate register usage. {\name} reduces registering by 45\% compared to Generic CGRA, \textcolor{black}{29\% compared to CGRA-Express-like} and 31\% compared to PreMap  by forwarding values within composed execution chains instead of writing them back at every PE boundary. Pre-Map and In-Map still insert register boundaries whenever compositions cannot be maintained across cycles, which increases the number of intermediate writes. This difference is more pronounced in workloads with long dependent chains, where {\name} can keep values within the combinational datapath across multiple operations. \textbf{{\name} reduces EDP by minimizing intermediate register accesses and unnecessary data movement.}

\textbf{Real-time Performance:} Fig~\ref{fig:edp} also shows the input-to-output latency across benchmarks. {\name} improves initiation interval and overall throughput, but this comes with a modest increase in pipeline depth, leading to slightly higher input-to-output latency compared to In-Map. In most workloads, this increase is limited to one additional pipeline stage, while in cases such as fft, the increase is more noticeable due to deeper dependency chains. In contrast, In-Map achieves lower input-to-output latency by maintaining a shallower pipeline, but at the cost of a higher initiation interval and reduced throughput. Pre-Map exhibits higher latency due to its coarse partitioning and is not competitive in this trade-off. 
\textcolor{black}{CGRA-Express-like baseline fuses only adjacent operations, so on kernels with longer chains it forms more pipeline stages than {\name}, giving higher input-to-output latency, while still failing to reduce the initiation interval on recurrence-bound kernels like dither.}
\textbf{This highlights a clear trade-off between throughput and latency. {\name} prioritizes throughput by reducing initiation interval through composition, while In-Map prioritizes latency by limiting pipeline depth.}


\subsection{Ablation Study}
\begin{figure}[h!]
	\centering
	\includegraphics[width=\columnwidth]{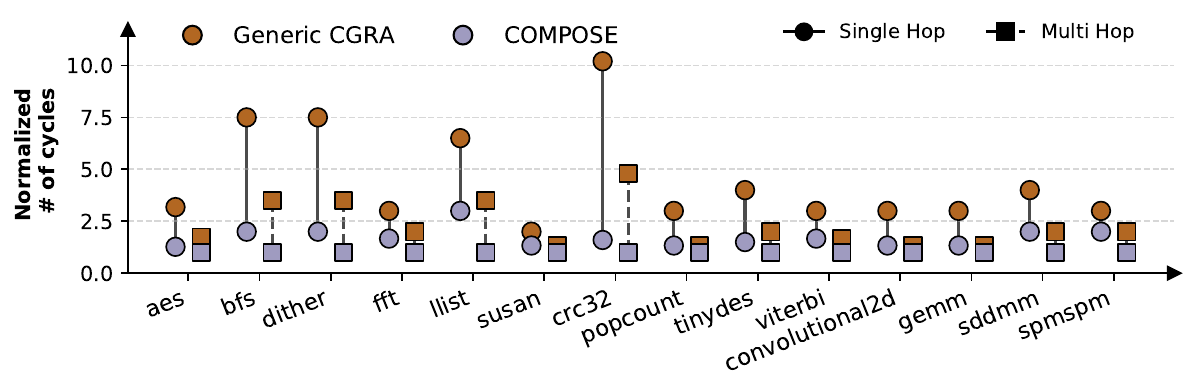}
    \vspace{-0.6cm}
	\caption{Cycle counts normalized to {\name} multi-hop across benchmarks. {\name} consistently achieves lower latency than Generic CGRA under both single-hop and multi-hop router communication.}
	\label{fig:hops}
\end{figure}
\textbf{Interconnect Model:} Fig.~\ref{fig:hops} compares {\name} across two fabrics: (i) single-cycle single-hop routing and (ii) single-cycle multi-hop routing, reporting performance (normalized cycle count) and the number of VPEs formed. In the multi-hop design, {\name} incorporates STA-derived router-to-router delay (crossbar delay) and scales it with hop count to determine whether a chain fits within a cycle. This directly affects VPE formation, allowing a single VPE to span multiple hops when timing permits. As a result, fewer but larger VPEs are formed, which improves cycle count by increasing the amount of work completed within each cycle.

In contrast, single-hop routing restricts VPE formation to adjacent PEs, forcing earlier VPE termination when dependencies span longer distances. This results in a higher number of smaller VPEs and increased pipeline boundaries, which leads to higher cycle counts. The difference is more pronounced in workloads such as bfs and fft, where dependencies are distributed across the fabric, while workloads like gemm show smaller differences due to localized mappings. \textbf{Overall, the reduction in VPE count in the multi-hop case directly correlates with improved performance, highlighting that flexible routing enables more effective composition.}

\begin{figure}[h!]
	\centering
	\includegraphics[width=\columnwidth]{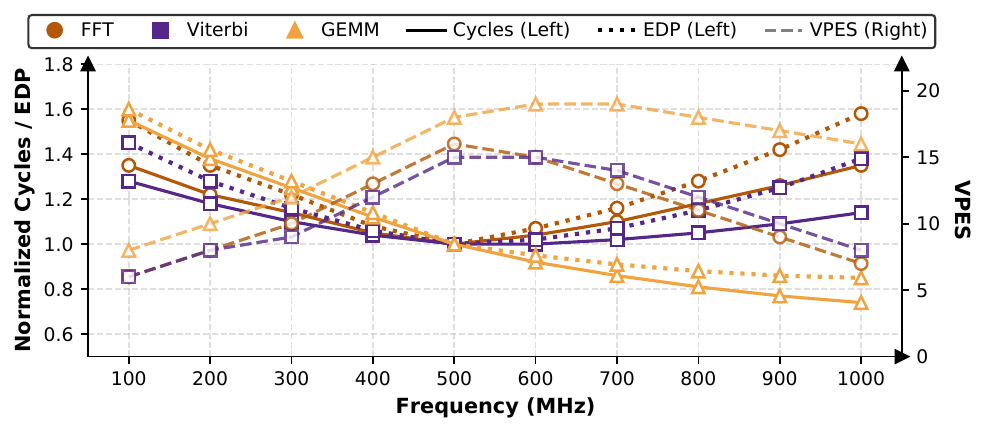}
    \vspace{-0.4cm}
	\caption{Impact of operating frequency on execution time, EDP, and VPE count across workloads.}
	\label{fig:freq_variation}
\end{figure}
\textbf{Frequency Variation:} Fig.~\ref{fig:freq_variation} shows the impact of operating frequency on normalized cycles, EDP, and VPE count for fft, viterbi, and gemm, which represent three classes of workloads. For fft and viterbi, the optimal operating point occurs around 500 MHz, where {\name} achieves the best trade-off between composition and timing. At this frequency, sufficient slack exists to form larger VPEs, enabling the mapper to reach the 
 achievable II. As frequency increases, the tighter $T_{clk}$ restricts VPE formation, reducing composition opportunities and increasing VPE count. This shifts the bottleneck to recurrence constraints in fft and limits further gains in viterbi. In contrast, gemm continues to benefit at higher frequencies since it is primarily resource-bound and less dependent on composition. \textcolor{black}{\textbf{Thus, the optimal frequency is not the highest clock frequency, but the point where {\name} maximizes VPE size while avoiding recurrence-limited execution, enabling peak performance at moderate frequencies.}}

\subsection{{\name} scales to large CGRAs}
\begin{figure}[h!]
	\centering
	\includegraphics[width=\columnwidth]{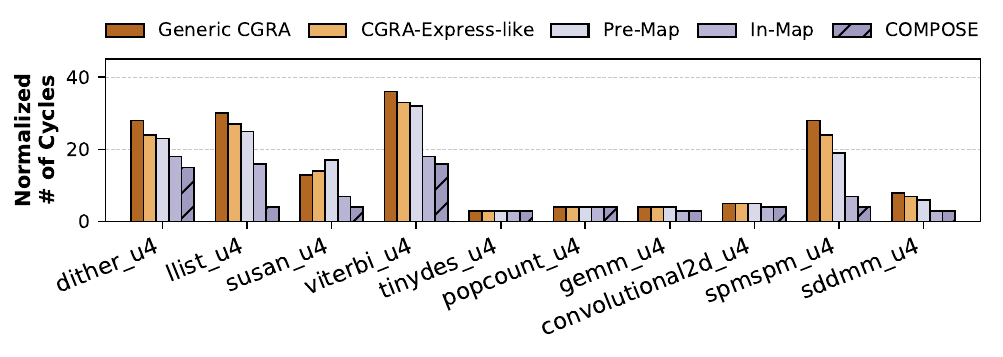}
    \vspace{-0.6cm}
	\caption{\textcolor{black}{Scalability on 8×8 CGRA for large DFGs; cycles normalized to {\name} show consistent improvements.}}
	\label{fig:scalability}
\end{figure}
Fig.~\ref{fig:scalability} evaluates {\name} on an $8\times8$ CGRA, showing that its benefits persist as the fabric size increases. {\name} continues to achieve lower cycle counts and improved EDP compared to all baselines, while maintaining efficient VPE formation across a larger spatial domain. The larger fabric provides more placement flexibility, enabling the mapper to form VPEs with longer chains without being constrained by local resource availability. 
\textcolor{black}{CGRA-Express-like baseline benefits less from this, as neighbor-only fusion cannot reach across the fabric to form longer chains.}
At the same time, {\name} effectively manages the increased routing complexity by incorporating hop-aware timing into VPE formation, ensuring that compositions remain timing-valid. \textbf{{\name} scales with CGRA size by converting increased spatial flexibility into larger VPEs and improved execution efficiency.}

\subsection{Overhead of {\name}}
{\name} introduces modest hardware overheads of 3.8\% in area and 2.3\% in power, arising from additional bypassing support in the PE datapath, specifically two input multiplexers and one predication multiplexer per PE, while output bypass paths already exist in generic CGRA designs. {\name} increases compilation time by approximately 1.6×, as VPE boundaries must be determined dynamically based on timing rather than fixed pipeline stages. \textbf{However, given the small hardware overhead and consistent performance gains, this trade-off is practical.} 

{\textcolor{black}{
\subsection{{\name} Generalizes Across Datatypes}
\begin{figure}[h!]
	\centering
	\includegraphics[width=\columnwidth]{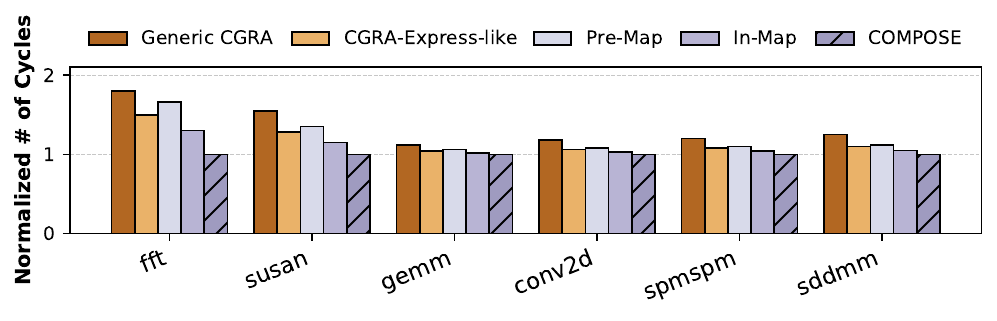}
    \vspace{-0.6cm}
	\caption{\textcolor{black}{FP16 cycle counts normalized to {\name}, with largest gains on recurrence-heavy kernels.}}
	\label{fig:datatype}
\end{figure}
Evaluation so far uses the integer datapath of our taped-out chip. To show {\name} is not tied to a datatype, we synthesized and placed-and-routed an FP16 ALU and extracted its per-operation delays through the same STA flow as Section~\ref{sec:sta_analysis} (post-PnR, not silicon-verified). Since {\name} consumes per-operation timing as an input, the framework is unchanged: the FP16 delays simply replace the integer $\delta(v)$ values during VPE formation.}

\textcolor{black}{Fig.~\ref{fig:datatype} reports FP16 cycle counts. {\name} retains up to a $1.7\times$ cycle reduction over Generic CGRA (on fft), smaller than the integer case because wider arithmetic has longer critical paths and leaves less slack to compose, most visible on the recurrence-heavy fft and susan. \textbf{Datatype only shifts the per-operation delays fed to VPE formation, so wider datapaths reduce composition opportunity without changing the framework or diminishing {\name}'s advantage over the baseline.}}
}

\section{Other Related Works}
Prior work can be broadly categorized based on the level at which they improve execution: (i) composability at the level of cores or accelerators, (ii) enhancements to PEs, and (iii) micro-architectural fusion and datapath pipelining. We discuss them in the context of composability and accelerating inter-iteration dependencies.

\textbf{Core-level composability.} Prior work has explored composability at coarser granularities, particularly at the level of processor cores or accelerator units. Stitch~\cite{stitch}, for example, proposes a many-core architecture where instruction set extension units are combined to form larger virtual accelerators. Similarly, \cite{burger_composable_multicore} enables runtime aggregation of small cores into more powerful logical processors to match application parallelism. While these approaches demonstrate the benefits of composability via compiler-scheduled NoCs and runtime reconfiguration, they operate at a coarse granularity and do not expose mechanisms for composing fine-grained compute elements as in a CGRA datapath.

\textbf{CGRAs with enhanced PEs.} Another approach to improve CGRA efficiency is increasing the functional richness of individual PEs while preserving a fixed abstraction. Plaid~\cite{plaid} introduces motif-based execution, mapping recurring communication patterns onto specialized units with three ALUs and a local router. Unlike Plaid’s fixed configuration, {\name} supports flexible virtual PE boundaries, enabling a variable number of ALUs. 
RF-CGRA~\cite{rf_cgra} enhances interconnects for multi-cycle dependencies using register chains, whereas {\name} improves efficiency for short logic paths by fusing operations and eliminating intermediate registers. FPCA~\cite{fully_pipelined} explores dynamic runtime composition for parallel workloads, which is complementary to {\name}’s compile-time approach. 
\textcolor{black}{NUPEA~\cite{nupea_isca2025} also starts from recurrence analysis but targets data movement, steering latency-critical loads into near-memory domains during place-and-route. {\name} instead uses recurrence information to drive composition; the two are complementary and could be combined.}

\textbf{Micro-architectural fusion and datapath pipelining.} Another line of work improves performance through micro-architectural optimizations and scheduling. 
Capstone~\cite{capstone} extends Cascade~\cite{cascade} with a compiler-driven energy model to enforce power constraints.
FLAME~\cite{flame} decomposes long-latency operations into multi-cycle execution for high-frequency operation. In contrast, {\name} targets lower-frequency regimes, achieving comparable or better performance by fusing operations into a combinational datapath. 
\textcolor{black}{More broadly, HLS harvests slack at synthesis time via cycle-time-aware scheduling and operation chaining~\cite{sivaraman02,ruizautua05}, with FPGA-targeted flows relying on estimated delay models~\cite{zaretsky07}; unlike a CGRA mapper, none expose measured post-layout STA delays to spatial placement on a fixed fabric or resolve the loop-carried recurrences that bound throughput.}
Beyond CGRAs, DPU-v2~\cite{dpu_v2} accelerates irregular DAGs by mapping them into tree-like structures aligned with its datapath.

Prior works optimize hardware and software in isolation, neglecting significant timing slack from CGRA operation variability. \textbf{{\name} bridges this divide through hardware-software co-design, exposing circuit timing directly to the compiler.}
\section{Discussion}
\textbf{Understanding {\name} through VLIW lens:}
While both {\name} and VLIW leverage instruction-level parallelism (ILP)~\cite{fisher_vliw} by grouping operations into execution packets, they differ in managing temporal constraints. In traditional VLIW, the compiler packs independent operations that execute in parallel across multiple functional units in lock-step. This synchronous execution often forces faster units to wait for the slowest operation in the bundle, creating significant timing slack. {\name} extends this concept through timing-driven spatial composition, allowing dependent producer-to-consumer operations to execute back-to-back within the same cycle. By utilizing STA to harvest slack from low-latency primitives and interconnect hops, {\name} reduces slack within the lock-step itself. This enables the framework to collapse multi-cycle recurrence paths and resolve complex dependencies in one cycle, avoiding the intermediate register file write-backs that standard VLIW architectures require for dependent chains.

\textbf{Understanding {\name}’s Chaining vs Cray-1's Vector Chaining~\cite{cray1}:}
While {\name} and the Cray-1 both utilize chaining, they differ fundamentally in their scheduling mechanisms and architectural intent. Cray-1's vector chaining is a runtime hardware process governed by a narrow chain slot time. If the dependent instruction is not ready to start at the exact moment the first result element exits the pipeline, the hardware misses the slot and forces the second instruction to wait for the entire vector to finish writing. In contrast, {\name}'s compilation strategy is entirely static and pre-calculates valid chaining opportunities before execution. While the Cray-1 chains operations across discrete clocked pipeline stages, {\name} chains operations within a single clock cycle.
\section{Conclusion}
This paper demonstrates that the rigid abstraction of fixed-cycle execution in CGRAs leaves significant performance gains and energy savings on the table. {\name} addresses this by exposing circuit-level timing to the compiler, 
allowing composable execution through dynamically formed virtual PEs. \textcolor{black}{This is especially beneficial for recurrence-bound loops where registered stages on such paths directly increase recurrence latency and can raise II. By 
collapsing recurrence-critical producer-consumer chains into fewer registered stages,} {\name} improves utilization, performance, energy, and resource efficiency, highlighting the value of tighter hardware-software co-design.



\bibliographystyle{ACM-Reference-Format}
\bibliography{ref}

\end{document}